\documentclass[a4paper,12pt]{article}

\usepackage{amssymb}

\begin{document}
\newcommand{\ds}{\displaystyle}
\newcommand{\be}{\begin{equation}}
\newcommand{\ee}{\end{equation}}
\newcommand{\ba}{\begin{array}}
\newcommand{\ea}{\end{array}}
\newcommand{\sn}{\mbox{sn}}
\newcommand{\cn}{\mbox{cn}}
\newcommand{\dn}{\mbox{dn}}
\newcommand{\bea}{\begin{eqnarray}}
\newcommand{\eea}{\end{eqnarray}}
\newcommand{\bi}{\begin{itemize}}
\newcommand{\ei}{\end{itemize}}
\newcommand{\x}{{\ensuremath{\times}}}
\newcommand{\bb}[1]{\makebox[16pt]{{\bf#1}}}
\newtheorem{theorem}{Theorem}
\newtheorem{definition}{Definition}
\newtheorem{lemma}{Lemma}
\newtheorem{comment}{Comment}
\newtheorem{corollary}{Corollary}
\newtheorem{example}{Example}
\newtheorem{examples}{Examples}
\newtheorem{acknowledgement}{Acknowledgement}
 
\title{ Nondegenerate 3D complex Euclidean superintegrable systems and algebraic varieties}
\author{E.\ G.\ Kalnins\\
{\sl Department of Mathematics,
 University
of Waikato,}\\{\sl Hamilton, New Zealand.}\\
J.~M.~Kress\\ 
{\sl School of Mathematics, The University of New South Wales,} \\
{\sl Sydney NSW 2052, Australia} \\
{\sl j.kress@unsw.edu.au} \\
and W.\ Miller, Jr.
\\
{\sl School of Mathematics, University of Minnesota,}\\
{\sl Minneapolis, Minnesota,
55455, U.S.A.}\\
{\sl miller@ima.umn.edu} \\
}
\date{\today}
\maketitle

\vspace{0.3cm}
%\tableofcontents

\begin{abstract} 
 A classical (or quantum)  second order superintegrable system  is an integrable
 $n$-dimensional Hamiltonian system with potential that 
admits $2n-1$ functionally independent second order constants of the motion
 polynomial in the momenta, the maximum possible. 
 Such systems have remarkable properties: multi-integrability and
 multi-separability, an  algebra of higher order symmetries whose 
representation theory yields spectral information about the
 Schr\"odinger operator, deep connections with  special functions and with QES
 systems.
  Here we announce a complete classification of nondegenerate (i.e., 4-parameter) potentials  for complex Euclidean 3-space. We  characterize the possible 
 superintegrable systems  as points on an algebraic variety in 10 variables subject to six
 quadratic polynomial constraints.  The Euclidean group
 acts on the variety such that  two points determine the same
 superintegrable system if and only if they lie on the same leaf of
 the foliation. There are exactly 10 nondegenerate potentials.

\end{abstract}

\section{Introduction} 
 
For any complex 3D  conformally flat  manifold we can always find local coordinates $x,y,z$ such that the classical  Hamiltonian takes the form
\be\label{hamiltonian}
H=\frac{1}{\lambda(x,y,z)}(p_1^2+p_2^2+p_3^2)+V(x,y,z),\qquad
(x,y,z)=(x_1,x_2,x_3),
\ee
i.e., the complex metric is 
$ds^2=\lambda(x,y,z)(dx^2+dy^2+dz^2)$.  
This system is  {\bf superintegrable} for some potential $V$ if it
admits 5 functionally independent constants of the motion (the maximum
number possible) that are
polynomials in the momenta $p_j$. It is {\bf second order
  superintegrable } if the constants of the motion are quadratic,
i.e., of the form 
\be\label{symmetry}
S=\sum a^{ji}(x,y)p_jp_i +W(x,y,z).\ee
 That is, $\{{ H},{S}\}=0$ where 
\[ \{f,g\}=\sum_{j=1}^n(\partial_{x_j}f\partial_{p_j}g-\partial_{p_j}f\partial_{x_j}g)
\]
is the Poisson bracket for functions $f({\bf x},{\bf p}),g({\bf
  x},{\bf p})$ on phase space \cite{WOJ,EVA,EVAN,FMSUW,FSUW,MSVW,CALO,CIMC}.
There is a similar definition of second order superintegrability
for quantum systems with formally self-adjoint Schr\"odinger and
symmetry operators whose classical analogs are those given above,
and these systems correspond one-to-one, \cite{KKM20061}. (In particular, the
  terms in the Hamiltonian that are quadratic in the momenta are
  replaced by the Laplace-Beltrami operator on the manifold, and  Poisson
  brackets are replaced by operator commutators in the quantum case.)  Historically the most
important superintegrable system is the Euclidean space Kepler-Coulomb
problem where  $V=\alpha/\sqrt{x^2+y^2+z^2}$. (Recall that this
system not only has angular momentum and energy as constants of the
motion but a Laplace vector that is conserved.) Second order superintegrable  systems have remarkable properties. In
particular, every trajectory of a solution of the Hamilton equations
for such a system in 6-dimensional phase space lies on the intersection
of 5 independent constant of the motion hypersurfaces in that space,
so that the trajectory can be obtained by algebraic methods alone,
with no need to solve Hamilton's equations directly. Other common properties
include multiseparability (which implies multiintegrability, i.e.,
integrability in distinct ways) \cite{WOJ,EVA,EVAN,FMSUW, FSUW,MSVW,
  CIMC,CALO,MPSTAN,GZLU,BDK}  and the existence of a quadratic algebra of symmetries that
closes at order 6. The quadratic algebra in the quantum case gives
information relating the spectra of the constants of the motion,
including the Schr\"odinger operator.

Many examples of 3D superintegrable  systems are known, although, in
distinction to the 2D case,  they
have not been classified, \cite{GPS, KKMP, KKW, KKMW, RAN, KMWP}.
Here,  we employ  theoretical methods based on integrability
conditions to obtain a complete classification  of Euclidean systems
with nondegenerate potentials.  To make it clear how these systems
relate to general second order superintegrable systems we  introduce
some terminology. A set of 2nd order symmetries  for a classical superintegrable system  is  either linearly independent (LI) or 
linearly dependent (LD). LI sets can   functionally independent (FI)
in the 6-dimensional phase space in two ways: they are strongly functionally independent  (FI-S) if they are
functionally independent even when the potential is set equal to zero. They are
weakly functionally independent (FI-W) if the functional independence
holds only when the potential is turned on (example: the isotropic
oscillator). Otherwise they are functionally dependent (FD). An LI set can be functionally linearly independent (FLD) if it is linearly dependent at each regular point, but the linear dependence varies with the point.   An LI set can be  FLD in two ways. It
is weakly functional linear dependent (FLD-W) if the functional linear
dependence holds only with the potential turned off and strongly
functional linear dependent (FLD-S) if the functional linear dependence
holds even with the potential turned on. Otherwise the set is
functionally linearly independent (FLI). The Calogero and Generalized Calogero
potentials are FD, and FLD-S, \cite{KKM20061}. One property of FLD systems is that their potentials satisfy a first order linear partial differential equation. Thus they can be expressed in terms of a function of only two variables. In that sense they are degenerate. This paper is concerned with  a classification of functionally linearly independent potentials. As shown in 
 \cite{KKM20051},  if a 3D second order superintegrable system is FLI then the potential  $V$ is  must satisfy a system of coupled PDEs of the form
\be \label{veqn1a} V_{22}=V_{11}+A^{22}V_1+B^{22}V_2+C^{22}V_3,\
V_{33}=V_{11}+A^{33}V_1+B^{33}V_2+C^{33}V_3,
\ee
$$
V_{12}= A^{12}V_1+B^{12}V_2+C^{12}V_3,\
V_{13}= A^{13}V_1+B^{13}V_2+C^{13}V_3,$$
\be\label{3Dnondegenerate} V_{23}= A^{23}V_1+B^{23}V_2+C^{23}V_3.\ee
The analytic functions $A^{ij},B^{ij},C^{ij}$ are determined uniquely from the Bertrand-Darboux equations for the 5 
constants of the motion and are analytic except for a finite number of poles. 
If the integrability conditions for these equations are satisfied identically then the potential is said to be {\bf nondegenerate}.  A nondegenerate potential (which is actually a vector space of potential functions) is characterized by the following property. At any regular point
${\bf x}_0=(x_0,y_0,z_0)$, i.e., a point where
the $A^{ij},B^{ij},C^{ij}$ are defined and analytic and the constants of the motion are functionally independent, 
 we can prescribe the values of
$V({\bf x}_0)$, $V_1({\bf x}_0)$,$V_2({\bf x}_0)$,$V_3({\bf x}_0)$,$V_{11}({\bf x}_0)$ arbitrarily and obtain a unique solution of (\ref{3Dnondegenerate}). Here,
$V_1=\partial V/\partial x$, $V_2=\partial V/\partial y$, etc. The 4 parameters for a nondegenerate potential (in addition to the usual additive constant) are the maximum number of parameters that can appear in a 
superintegrable system. A FLI superintegrable system is {\bf degenerate} if the potential function satisfies additional restrictions in addition to equations  (\ref{3Dnondegenerate}).  These restrictions can arise in two ways, either as additional equations arising directly from the Bertrand-Darboux equations or as restrictions that occur because the integrability conditions for equations (\ref{3Dnondegenerate}) are not satisfied identically. In any case, the number of free  parameters for a degenerate potential is strictly fewer than 4. In this sense the nondegenerate potentials are those of maximal symmetry, though the symmetry is not meant in the traditional Lie group or Lie algebra  sense. Nondegenerate  potentials admit no nontrivial Killing vectors. 
Our concern in this paper is the classification of all 3D FLI
nondegenerate potentials in complex Euclidean space. In \cite{KKM20071} we
have begun the study of fine structure for second order 3D
superintegrable systems, i.e., the structure and classification theory
of systems with various types of degenerate potentials.

Our plan of attack is as follows. First we give a brief
review of the fundamental equations that characterize second order
FLI systems with nondegenerate potential in a 3D conformally flat
space. Then we review the structure theory that has been worked out
for these systems, including multiseparability and the existence of a
quadratic algebra.  We will recall the fact that all  such systems are
equivalent via a St\"ackel transform to a superintegrable system on
complex Euclidean 3-space or on the complex 3-sphere. Thus a
classification theory must focus on these two spaces. Due to the
multiseparability of these systems we can use separation of variables
theory to help attack the classification problem. In \cite{KKM20052} we showed
that associated with each of the 7 Jacobi elliptic coordinate generically
separable systems for complex Euclidean space there was  a
unique superintegrable system with a separable  eigenbasis in these
coordinates. Thus the only remaining systems were those that separated
in nongeneric orthogonal coordinates alone, e.g., Cartesian coordinates, spherical
coordinates, etc. The possible nongeneric separable coordinates are
known \cite{ERNIE} so, in principle, the classification problem could be solved.    Unfortunately, that still left so many specific
coordinate systems to check that classification was a practical
impossibility. Here we present a new attack on the problem, based on 
 characterizing  the possible 
 superintegrable systems with nondegenerate potentials as points on an algebraic variety.
 Specifically, we determine a variety in 10 variables subject to six
 quadratic polynomial constraints. Each point on the
 variety corresponds to a superintegrable system. The Euclidean group
 $E(3,\bb C)$ acts on the variety such that  two points determine the same
 superintegrable system if and only if they lie on the same leaf of
 the foliation. The differential equations describing the spacial
 evolution of the system are just those induced by the Lie algebra of
 the subgroup of Euclidean translations. A further simplification is
 achieved by writing the algebraic and differential equations in an
 explicit 
 form so that they transform irreducibly according to representations
 of the rotation subgroup $SO(3,\bb C)$. At this point the equations
 are simple enough to check directly which superintegrable systems
 arise that permit separation in a given coordinate system.  We show that in addition to the 7 superintegrable systems corresponding to separation in one of the generic separable coordinates, there are exactly 3 superintegrable systems that separate only in nongeneric coordinates. Furthermore, for every system of orthogonal separable coordinates in complex Euclidean space there corresponds at least one nondegenerate superintegrable system that separates in these coordinates. The method of proof of these results should generalize to higher dimensions.

\section{Conformally flat spaces in three dimensions} Here we review some basic results about 3D second order superintegrable systems in conformally flat spaces. 
 For each such space there always exists a local coordinate system 
  $x,y,z$ and a nonzero function $\lambda(x,y,z)=\exp G(x,y,z)$ such that the
 Hamiltonian is (\ref{hamiltonian}).
A
 quadratic constant of the motion   (or generalized symmetry) (\ref{symmetry})
must satisfy   
$
\{{ H},{\ S}\}=0
$.
i.e.,
\begin{equation}\label{killingtensors}\begin{array}{lll}
a_i^{ii}&=&-G_1a^{1i}-G_2a^{2i}-G_3a^{3i}\nonumber\\
2a^{ij}_i+
a_j^{ii}&=&-G_1a^{1j}-G_2a^{2j}-G_3a^{3j},\quad i\ne j\nonumber\\
 a^{ij}_k+a^{ki}_j+a^{jk}_i&=&0,\quad i,j,k\ {\rm distinct}\noindent\end{array}
\end{equation}
and
\begin{equation}
W_k=\lambda\sum_{s=1}^3a^{sk}
V_s,\quad k=1,2,3
.\label{potc}
\end{equation}
(Here a subscript $j$ denotes differentiation with respect to $x_j$.)
The requirement that $\partial_{x_\ell} W_j=\partial_{x_j}
W_\ell,\ \ell\ne j$ leads from
(\ref{potc}) to the
second order Bertrand-Darboux  partial differential equations for the potential.
\begin{equation}\label{BertrandDarboux}
\sum_{s=1}^3\left[V_{sj}\lambda a^{s\ell}-V_{s\ell}\lambda a^{sj}+
V_s\left(
(\lambda a^{s\ell})_j-(\lambda
a^{sj})_\ell\right)\right]=0.
\end{equation}

  For second order superintegrabilty in 3D there must
be five functionally independent constants of the motion (including the Hamiltonian itself). Thus  the Hamilton-Jacobi 
equation admits four   additional constants of the 
motion: 
\[ %\label{3dconst}
{ S}_h=\sum_{j,k=1}^3a^{jk}_{(h)}p_kp_j+W_{(h)}={ L}_h+W_{(h)},\qquad h=1,\cdots,4.
\]
We assume that the four functions ${ { S}}_h$ together with ${H}$ are functionally linearly
independent in the six-dimensional phase space. In \cite{KKM20051} it is shown that the matrix of the 15 B-D equations for the potential  has rank at least 5, hence we can solve for the second derivatives of the potential in the form (\ref{veqn1a}).
If the matrix  has rank $>5$ then there will be
additional conditions on the potential and it will depend on fewer parameters.
$D^{1}_{(s)}V_1+D^{2}_{(s)}V_2+D^{3}_{(s)}V_3=0$. Here the $A^{ij},B^{ij},C^{ij},D^{i}_{(s)}$ are
functions of $x$, symmetric in the superscripts, that can be calculated explicitly. 
Suppose now that the superintegrable system is such that the rank is exactly 5 so that the relations are only 
 (\ref{veqn1a}). Further, suppose the integrability conditions for system  (\ref{veqn1a})  are satisfied identically. 
In this case  the potential is  nondegenerate. 
Thus,  at any point ${\bf x}_0$, where the $A^{ij}, B^{ij}, C^{ij}$ are 
defined and analytic, there is a unique solution $V({\bf x})$ with
arbitrarily prescribed values of $V_1({\bf x}_0), V_2({\bf x}_0),V_3({\bf x}_0),V_{11}({\bf x}_0)$ (as well as the value of $V({\bf x}_0)$ itself.)
The points ${\bf x}_0$ are called {\it regular}. 

Assuming that $V$ is nondegenerate, we
 substitute the requirement  (\ref{veqn1a}) into the
B-D equations (\ref{BertrandDarboux}) and obtain three equations for the
derivatives $a^{jk}_i$.
Then we can  equate coefficients of $V_1,V_2,V_3,V_{11}$ on each side of the  conditions $\partial_1V_{23}=\partial_2V_{13}
=\partial_3V_{12}$, $\partial_3V_{23}=\partial_2V_{33}$, etc., to obtain
   integrability conditions, the simplest of which are
\begin{equation}\label{int11}
A^{23}=B^{13}=C^{12},\ B^{12}-A^{22}=C^{13}-A^{33},\
B^{23}=A^{31}+C^{22},\ C^{23}=A^{12}+B^{33}.
\end{equation} 
It follows that the 15 unknowns can be expressed linearly in terms of the 10
functions 
\be\label{10terms} A^{i2},A^{13},A^{22},A^{23},A^{33}, B^{12}, B^{22},B^{23},B^{33}, C^{33}.\ee
In general, the integrability conditions satisfied by the potential
equations take the following form. We  introduce  the vector
${\bf w}=\left( V_1, V_2, V_3,
V_{11}\right)^{\rm T}$,
and the matrices
${\bf A}^{(j)}$, $j=1,2,3$, such that
\be\label{int21}
\partial_{x_j}{\bf w}={\bf A}^{(j )}{\bf w}\qquad  j=1,2,3.
\ee
The integrability conditions for this system are
\be\label{int31}
A^{(j)}_i-A^{(i)}_j=A^{(i)}A^{(j)}-A^{(j)}A^{(i)}\equiv [A^{(i)},A^{(j)}].
\ee
The integrability conditions (\ref{int11}) and (\ref{int31}) are analytic expressions in $x_1,x_2,x_3$ and must hold identically. 
Then the system has a solution $V$ depending on 4 parameters (plus an arbitrary additive parameter).

Using the nondegenerate potential condition and the B-D equations  we can solve for
all of the first partial derivatives $a^{jk}_i$  of a  quadratic
symmetry to obtain the 18 basic symmetry equations, (27) in \cite{KKM20051}, 
plus the linear relations (\ref{int11}).
Using the  linear relations we can express
$C^{12},C^{13},C^{22},C^{23}$ and $B^{13}$ in terms of the remaining
$10$ functions. Each $a^{jk}_i$ is a linear combination of the
$a^{\ell m}$ with coefficients that are linear in the 10 variables
and in the $G_s$.

Since this system of first order partial differential equations
is involutive, the general
solution for the 6 functions $a^{jk}$ can depend on at most 6
parameters, the values $a^{jk}({\bf x}_0)$ at a fixed regular point
${\bf x}_0$. For the  integrability conditions
 we define the vector-valued function
\[
{\bf h}(x,y,z)=\left(
  a^{11},\\a^{12},\\a^{13},\\a^{22},\\a^{23},\\a^{33}\right)^{\rm T}
\]
and  directly compute the $6\times 6$ matrix functions ${\cal A}^{(j)}$ to get the first-order system
$ %\label{int4c}
\partial_{x_j}{\bf h}={\cal A}^{(j )}{\bf h},$ $  j=1,2,3$.
The integrability conditions for this system are are
\begin{equation}\label{int5c}
{\cal A}^{(j)}_i{\bf h}-{\cal A}^{(i)}_j{\bf h}={\cal A}^{(i)}{\cal A}^{(j)}{\bf h}-{\cal A}^{(j)}{\cal A}^{(i)}{\bf h}\equiv [{\cal A}^{(i)},{\cal A}^{(j)}]{\bf h}.
\end{equation}

By assumption we have 
 5 functionally linearly independent symmetries, so at each regular
 point the solutions sweep out a 5 dimensional subspace of the 6
 dimensional space of symmetric matrices. However, from the conditions
 derived above there seems to be no obstruction to construction of a 6
 dimensional space of solutions. Indeed in \cite{KKM20051} we show that this
 construction can always be carried out.

\begin{theorem} $ (5\Longrightarrow 6)$  Let $V$ be a nondegenerate potential corresponding to
  a conformally flat space in 3 dimensions that is superintegrable,
i.e., suppose $V$ satisfies the equations (\ref{veqn1a}) whose
integrability conditions
 hold identically, and there are 5 functionally
independent constants of the motion.  Then the space of second order symmetries
 for the Hamiltonian ${ H}=(p^2_x+p^2_y+p^2_z)/\lambda(x,y,z)+V(x,y,z)$
(excluding multiplication by a constant) is of
dimension $D= 6$. 
\end{theorem}

Thus, at any regular point $(x_0,y_0,z_0)$, and given constants
$\alpha^{kj}=\alpha^{jk}$, there is exactly one symmetry  ${ S}$ (up to an additive constant) such
that $a^{kj}(x_0,y_0,z_0)=\alpha^{kj}$. Given a set of $5$ functionally
independent 2nd order  symmetries ${\cal L}=\{{ S}_\ell:\ell=1,\cdots 5\}$ associated with the
potential, there is always a $6$th second order symmetry ${S}_6$ that is
functionally dependent on $\cal L$, but linearly independent. 

Since the solution space of the symmetry equations  is of
dimension $D=6$, it follows that the integability conditions for these
equations must be satisfied identically in the $a^{ij}$
As part of the analysis in reference \cite{KKM20051} we  used the integrability
conditions for these equations and for the  
potential to derive the following:

\begin{enumerate} \item
An expression for each of the first partial derivatives $\partial_\ell A^{ij}$,  $\partial_\ell B^{ij}$, $\partial_\ell C^{ij}$, 
for the $10$ independent functions
 as homogeneous polynomials
of order at most two in the $A^{i'j'}$, $B^{i'j'}$,  $C^{i'j'}$. There are
 $30=3\times 10$ such expressions in all.
 (In the case $G\equiv 0$ the full set of conditions can be written in
the convenient form (\ref{Zde}), (\ref{Yde}).)
\item Exactly  5 quadratic identities for the
$10$ independent functions, see (31) in \cite{KKM20051}. In Euclidean space
  these identities take the form $I^{(a)} - I^{(e)}$ in (\ref{ideal})
  of the present paper.
\end{enumerate}
In references \cite{KKM20051} we studied the structure of the spaces of third, fourth and sixth order symmetries (or constants of the motion) of $H$. Here the {\bf order} refers to the highest order terms in the momenta. We  established the following results. 
\begin{theorem} Let $V$ be a superintegrable nondegenerate
  potential on a conformally flat space. Then
  the space of  third order constants of the motion is 4-dimensional
  and is spanned by Poisson brackets $R_{jk}=\{S_j,S_k\}$ of the second order constants of
  the motion. The 
 dimension of the space of  fourth order symmetries  is $21$ and is spanned by second order polynomials in the 6 basis symmetries $S_h$. (In particular, the Poisson brackets $ \{R_{jk},S_\ell\}$ can be expressed as second order polynomials in the basis symmetries.)
The  dimension of the space of 
  sixth order symmetries is  $56$ and is spanned by third order polynomials in the 6 basis symmetries $S_h$. (In particular the products $R_{jk}R_{\ell h}$ can be expressed  by third order polynomials in the 6 basis symmetries.) 
\end{theorem}
There is a similar result for fifth order constants of the motion, but it  follows directly from the Jacobi identity for the Poisson bracket. This establishes the quadratic algebra structure of the space of constants of the motion: it is closed under the Poisson bracket action. 

{}From the general theory of variable separation for Hamilton-Jacobi
equations  \cite{ERNIE, MIL88} and  the structure theory for Poisson brackets of second order constants of the motion, we established the following result \cite{KKM20052}.
\begin{theorem}\label{3Dmultiseparable} A  superintegrable system with  nondegenerate
  potential in a 3D conformally flat  space is 
  multiseparable. That is, the Hamilton - Jacobi equation for the system can be solved by additive separation of variables in more than one orthogonal coordinate system.
\end{theorem}
The corresponding Schr\"odinger eigenvalue equation for the quantum systems can be solved by multiplicative separation of variables in the same coordinate systems.

Finally, in \cite{KKM20052} we studied  the St\"ackel transform for 3D systems,  
 an invertible  transform that maps a nondegenerate superintegrable system on one conformally flat manifold to a nondegenerate superintegrable system on another manifold. Our principal result was 
\begin{theorem}
Every superintegrable system with nondegenerate potential on a 3D conformally flat space is equivalent under the St\"ackel  transform to a superintegrable system on either 3D flat space or the 3-sphere.
\end{theorem}

\section{Generic separable coordinates for Euclidean spaces}
Now we turn to the classification of second order nondegenerate superintegrable systems in 3D complex Euclidean space.  A subclass of these systems can be obtained rather easily from separation of variables theory. To make this clear  we  recall some facts about generic 
elliptical coordinates in complex Euclidean $n$ space  and their
relationship to superintegrable systems with nondegenerate potentials
(see \cite{KKWMPOG} for more details).

Consider a second order superintegrable   system of the form 
 $H=\sum_{k=1}^n p_k^2+V({\bf x})$ 
in Euclidean $n$ space, expressed in Cartesian coordinates $x_k$. In
 analogy with the 3D theory, the potential is  nondegenerate if it satisfies a system of equations of the form
\be\label{nondegenerate} V_{jj}-V_{11} = \sum_{\ell=1}^nA^{jj,\ell}({\bf x})V_\ell,\quad j=2,\cdots ,n,\ee
$$ V_{kj}= \sum_{\ell=1}^nA^{kj,\ell}({\bf x})V_\ell,\quad 1\le  k<j\le n,$$
where all of the integrability conditions for this system of partial
differential equations are identically satisfied, \cite{KKM20041,KKM20051}. 
There is an important subclass of such nondegenerate superintegrable systems that can be constructed for all $n\ge 2$, based on their relationship to variable separation in generic Jacobi elliptic coordinates.
The
prototype superintegrable system which is nondegenerate in $n$ dimensional flat
space has the Hamiltonian 
\be \label{prototype}
 H=\sum ^n_{i=1}(p^2_i+ \alpha x^2_i + \frac{\beta _i}{ x^2_i} )+\delta.
\ee
This system  is  superintegrable with nondegenerate potential and a basis of 
$n(n+1)/2$ second order symmetry operators  given by 
$$P_i=p^2_i+\alpha x^2_i+ \frac{\beta _i}{ x^2_i},\quad
J_{ij}=(x_ip_j-x_jp_i)^2+\beta _i \frac{x^2_j}{ x^2_i} + \beta _j 
\frac{x^2_i}{ x^2_j},\quad i\neq j.
$$
Although there appear to be ``too many" symmetries, all are functionally dependent on a subset of $2n-1$ functionally independent symmetries.  A crucial
observation is that the corresponding  Hamilton-Jacobi equation $H=E$ admits additive 
separation in $n$ generic  elliptical coordinates. 
$$x^2_i=c^2 {\Pi ^n_{j=1}(u_j-e_i)}/{ \Pi _{k\neq i}(e_k-e_i)}
$$
simultaneously {\it for all}   values of the parameters with  $e_i\neq e_j$ if $i\neq j$ and $i,j=1,\cdots,n$. (Similarly the quantum problem $H\Psi = E\Psi$ is superintergrable and admits multiplicative separation.)  Thus the equation is multiseparable
and separates in a continuum of elliptic coordinate systems (and in
many others besides). The $n$ involutive  symmetries characterizing a fixed elliptic separable system are polynomial functions of the $e_i$, and requiring separation for all $e_i$ simultaneously sweeps out the full $n(n+1)/2$ space of symmetries and uniquely determines the nondegenerate potential.   The infinitesimal distance in Jacobi 
elliptical coordinates $u_j$ has the form 
\be\label{ellipticmetric}ds^2=-\frac{c^2}{ 4} \sum ^n_{i=1} 
\frac{\Pi _{j\neq i}(u_i-u_j)}{ \Pi ^n_{k=1}(u_i-e_k)} du^2_i
=-{c^2\over 4} \sum ^n_{i=1} \frac{\Pi _{j\neq i}(u_i-u_j)}{ P(u_i)}
du^2_i, 
\ee
where $P(\lambda )=\Pi ^n_{k=1}(\lambda -e_k)$. However, it is well
known that (\ref{ellipticmetric}) is a flat space metric for any
polynomial $P(\lambda)$ of order $\le n$ and that each choice of such
a $P(\lambda)$ defines an elliptic type multiplicative separable
solution of  the Laplace - Beltrami eigenvalue problem (with constant
potential) in complex Euclidean $n$-space, \cite{ERNIE}. The distinct cases are
labeled by the degree of the polynomial and the multiplicities of  its
distinct roots. If for each distinct case we determine the most
general potential that admits separation for all $e_i$ compatible with
the multiplicity structure of the roots, we obtain a unique
superintegrable system with nondegenerate potential and $n(n+1)/2$
second order symmetries, \cite{KKWMPOG, KKM20052}. These are the generic superintegrable systems.  (Thus, for $n=3$  there are 7 distinct cases for $-\frac14\ P(\lambda)$:
$$ (\lambda-e_1)(\lambda-e_2)(\lambda-e_3),\     (\lambda-e_1)(\lambda-e_2)^2,\  (\lambda-e_1)^3,\  $$
$$ (\lambda-e_1)(\lambda-e_2),\     (\lambda-e_1)^2,\  (\lambda-e_1),\  1,$$
where $e_i\ne e_j$ for $i\ne j$. The first case corresponds to Jacobi elliptic coordinates.) 
The number of distinct generic superintegrable systems for each integer $n\ge 2$ is 
$\sum_{j=0}^np(j)
$,
where $p(j)$ is the number of integer partitions of $j$. 

All of the generic separable systems, their potentials and their
defining symmetries can be obtained from the basic Jacobi elliptic
system in $n$  dimensions  by a complicated but well defined set of
limit processes \cite{ KKM20052, KKWMPOG,Bocher}. In addition to these generic superintegrable systems there is an undetermined number of nongeneric systems. For $n=2$ all the systems have been found, and now we give the results for $n=3$.

We review some of the details from reference  \cite{KKM20052} to show how  each of
the generic separable systems in three dimensions  uniquely determines a nondegenerate
superintegrable system that contains it. 
We begin by
summarizing the full list of orthogonal  separable systems in complex Euclidean
space and the associated symmetries. (All of these systems have been classified, \cite{ERNIE}, and all can be obtained from the ultimate generic 
 Jacobi elliptic coordinates  by limiting processes \cite{Bocher, KMR}.)   Here, a ``natural'' basis
for first order symmetries (Killing vectors) is given by
$ p_1\equiv p_x$, $ p_2\equiv p_y$,$  p_3\equiv p_z$, $
 J_1=yp_z-zp_y$,$ J_2=zp_x-xp_z$, $J_3=xp_y-yp_x$
in the classical case and 
$ p_1=\partial_x$, $ p_2=\partial_y$, $  p_3=\partial_z$, $
 J_1=y\partial_z-z\partial_y$, $ J_2=z\partial_x-x\partial_z$, $ J_3=x\partial_y-y\partial_x$
in the quantum case. (In the operator characterizations for the
quantum case, the classical product of two constants of the motion is replaced 
by the symmetrized product of the corresponding operator symmetries.)
The free Hamiltonian is $H_0=p_1^2+p_2^2+p_3^2$.  In each case below
we list the coordinates. The constants of the motion that characterize
these coordinates can be found in \cite{KKM20052}. We use the bracket
notation of Bocher \cite{Bocher} to characterize each separable system.

$$ [2111]\quad
x^2 =c^2 {(u-e_1)(v-e_1)(w-e_1)\over (e_1-e_2)(e_1-e_3)},\quad 
y^2 =c^2 {(u-e_2)(v-e_2)(w-e_2)\over (e_2-e_1)(e_2-e_3)}$$
$$z^2 =c^2 {(u-e_3)(v-e_3)(w-e_3)\over (e_3-e_1)(e_3-e_2)}$$
$$ [221]\quad
x^2+y^2=-c^2\left[\frac{(u-e_1)(v-e_1)(w-e_1)}{(e_1-e_2)^2}\right]
$$
$$-\frac{c^2}{e_1-e_2}\left[
 (u-e_1)(v-e_1)+(u-e_1)(w-e_1)+(v-e_1)(w-e_1)\right],
$$
$$ (x-iy)^2=c^2\frac{(u-e_1)(v-e_1)(w-e_1)}{e_1-e_2},\quad z^2=c^2\frac{(u-e_2)(v-e_2)(w-e_2)}{(e_2-e_1)^2}.$$
$$ [23] \quad
x-iy={1\over 2}c( {u^2+v^2+w^2\over uvw}- {1\over 2} 
{u^2v^2+u^2w^2+v^2w^2\over u^3v^3w^3}),$$ 
$$z={1\over 2}c({uv\over w} + {uw\over v} + {vw\over u}),\quad
x+iy=cuvw.
$$
$$ [311] \quad
x={c\over 4}(u^2+v^2+w^2+{1\over u^2}+{1\over v^2}+{1\over w^2})+{3\over 2}c,$$ $$y=-{c\over 4} {(u^2-1)(v^2-1)(w^2-1)\over uvw},\quad 
z=i{c\over 4} {(u^2+1)(v^2+1)(w^2+1)\over uvw}.$$
$$ [32] \quad
x+iy=uvw,\quad x-iy=-({uv\over w}+{uw\over v}+{vw\over u}),\quad 
z={1\over 2}(u^2+v^2+w^2).$$
$$ [41] \quad
x+iy=u^2v^2+u^2w^2+v^2w^2-{1\over 2}(u^4+v^4+w^4),\ 
x-iy=c^2(u^2+v^2+w^2),\ 
z=2icuvw.$$
$$[5] \quad
x+iy=c(u+v+w),\quad  
x-iy={c\over 4}(u-v-w)(u+v-w)(u+w-v),$$ 
$$z=-{c\over 4}(u^2+v^2+w^2-2(uv+uw+vw)).$$

\noindent We summarize the remaining degenerate separable coordinates:

{\bf Cylindrical type  coordinates.} All of these have one symmetry  in common:
$L_1=p^2_3.$ The 7 systems are, polar, Cartesian, light cone, elliptic,
parabolic, hyperbolic and semihyperbolic.

{\bf Complex sphere coordinates.}
 These all have the symmetry  
$L_1=J^2_1+J^2_2+J^2_3$ in common. The 5 systems are spherical, horospherical,
 elliptical, hyperbolic, and semi-circular parabolic.

{\bf Rotational types of coordinates.} There are 3 of these systems,
each of which is characterized by the fact that the momentum terms in  one defining symmetry
form a perfect square whereas the other two are not squares.

In addition to these orthogonal coordinates, there is a class of
nonorthogonal heat-type separable coordinates   that are related to
the embedding of the heat equation  in two dimensions into three
dimensional complex Euclidean space.\cite{ERNIE}. These coordinates are not present in real Euclidean space,  only in real Minkowski spaces.The coordinates do not have any bearing on our further analysis as they do not occur in nondegenerate systems in three dimensions. This is because they  are characterized by an element of the Lie algebra $p_1+ip_2$ (not squared, i.e., a  Killing vector) so they cannot occur for a nondegenerate system. 

Note that the first $7$ separable systems are ``generic,'' i.e.,
they occur in one, two or three - parameter families, whereas the
remaining systems are special limiting cases of the generic ones. Each of the $7$ generic Euclidean separable systems depends on a
 scaling parameter $c$ and up to three parameters $e_1,e_2,e_3$. For
 each such set of coordinates there is exactly one nondegenerate
 superintegrable system that admits separation in these coordinates
 {\it simultaneously for all values of the parameters $c,e_j$}. 
Consider the system $[23]$, for example. If a nondegenerate
superintegrable system separates in these coordinates for all values of
the parameter $c$, then the space of second order symmetries must
contain the $5$ symmetries 
$$ { H}=p_x^2+p_y^2+p_z^2+V,\quad { S}_1=J_1^2+J_2^2+J_3^2+f_1,\quad
{ S}_2=J_3(J_1+iJ_2)+f_2,$$
$${ S}_3=(p_x+ip_y)^2+f_3,\quad {S}_4=p_z(p_x+ip_y)+f_4.$$
It is straightforward to check that the $12\times 5$ matrix of coefficients of the
second derivative terms in the $12$ Bertrand-Darboux equations
associated with symmetries ${S}_1,\cdots, { S}_4$ has rank 5
in general. Thus, there is at most one nondegenerate superintegrable
system admitting these symmetries. Solving the Bertrand-Darboux
equations for the potential we find  the unique solution
$$
V({\bf x}):=\alpha(x^2+y^2+z^2)+\frac{\beta}{(x+iy)^2}+\frac{\gamma
  z}{(x+iy)^3}+\frac{\delta (x^2+y^2-3z^2)}{(x+iy)^4}.$$
Finally, we can use the symmetry conditions for this potential to
obtain the full $6$-dimensional space of second order
symmetries. This is the superintegrable system III on the following table. The other six cases yield corresponding results.

\begin{theorem} Each of the $7$ ``generic'' Euclidean separable
  systems determines a unique nondegenerate superintegrable system
 that permits separation simultaneously for all values of the 
 scaling parameter $c$ and any other defining parameters $e_j$. For
 each of these systems there is a basis of $5$ (strongly) functionally
 independent and $6$ linearly independent second order symmetries.
 The corresponding nondegenerate potentials and basis of symmetries are:
\be {\bf \rm  I\ } [2111] \qquad V={\alpha _1\over x^2} + {\alpha _2\over y^2} + {\alpha _3\over z^2} 
+\delta (x^2+y^2+z^2),
\ee
$$
{\cal P}_i=p ^2_{x_i}+\delta x^2_i+ {\alpha _i\over x^2_i},\qquad
{\cal J}_{ij}=(x_ip_{x_j}-x_jp_{x_i})^2+\alpha ^2_i {x^2_j\over x^2_i} + \alpha ^2_j 
{x^2_i\over x^2_j},\quad i\geq j. 
$$
\be {\bf\rm II\ }  [221] \qquad
 V=\alpha (x^2+y^2+z^2)+ \beta  {x-iy\over (x+iy)^3} +  {\gamma \over (x+iy)^2}
+ {\delta \over z^2},
\ee
$${\cal S}_1=J\cdot J+f_1,\quad {\cal S}_2=p^2_z+f_2,\quad {\cal S}_3=J^2_3+f_3,
$$
$$ 
{\cal S}_4=(p_x+ip_y)^2+f_4,\quad L_5=(J_2-iJ_1)^2+f_5.
$$
\be {\bf\rm  III\   }[23] \qquad
 V=\alpha (x^2+y^2+z^2)+  {\beta \over (x+iy)^2} + {\gamma z\over (x+iy)^3}+ 
{\delta (x^2+y^2-3z^2)\over (x+iy)^4},
\ee
$$
{\cal S}_1=J\cdot J+f_1,\quad {\cal S}_2=(J_2-iJ_1)^2+f_2,\quad {\cal S}_3=J_3(J_2-iJ_1)+f_3,
$$
$$
{\cal S}_4=(p_x+ip_y)^2+f_4,\quad  {\cal S}_5=p_z(p_x+ip_y)+f_5.
$$
\be {\bf \rm  IV\ }  [311]   \qquad 
 V=\alpha (4x^2+y^2+z^2)+ \beta x +{\gamma \over y^2} + {\delta \over z^2},
\ee
$${\cal S}_1=p^2_x+f_1,\quad {\cal S}_2=p^2_y+f_2,\quad {\cal S}_3=p_zJ_2+f_3,
$$
$$ {\cal S}_4=p_yJ_3+f_4,\quad {\cal S}_5=J^2_1+f_5.
$$
\be {\bf\rm  V\ }  [32]     \qquad
 V=\alpha (4x^2+y^2+z^2)+\beta x+{\gamma \over (y+iz)^2} + 
{\delta (y-iz)\over (y+iz)^3},
\ee
$$ {\cal S}_1=p^2_x+f_1,\quad {\cal S}_2=J^2_1+f_2,\quad {\cal S}_3=(p_z-ip_y)(J_2+iJ_3)+f_3,
$$
$$ {\cal S}_4=p_zJ_2-p_yJ_3+f_4,\quad {\cal S}_5=(p_z-ip_y)^2+f_5.
$$
\be {\bf\rm VI\ }  [41]   \
 V=\alpha \left(z^2-2(x-iy)^3+4(x^2+y^2)\right)+\beta \left(2(x+iy)-3(x-iy)^2\right)+\gamma (x-iy)+
{\delta \over z^2},
\ee
$${\cal S}_1=(p_x-ip_y)^2+f_1,\quad {\cal S}_2=p^2_z+f_2,\quad  {\cal S}_3=p_z(J_2+iJ_1)+f_3,
$$
$$ {\cal S}_4=J_3(p_x-ip_y)-{i\over 4}(p_x+ip_y)^2+f_4,\quad {\cal S}_5=(J_2+iJ_1)^2+4ip_zJ_1+f_5.
$$
\be{\bf\rm VII\ }  [5]    \qquad
 V=\alpha (x+iy)+\beta (\frac{3}{ 4}(x+iy)^2+\frac{1}{ 4}z)+\gamma ((x+iy)^3+ 
\frac{1}{ 16}(x-iy)+\frac{3}{ 4}(x+iy)z)
\ee
$$
+\delta (\frac{5}{ 16}(x+iy)^4+\frac{1}{ 16}(x^2+y^2+z^2)+
\frac{3}{ 8}(x+iy)^2z),
$$
$${\cal S}_1=(J_1+iJ_2)^2+2iJ_1(p_x+ip_y)-J_2(p_x+ip_y)+\frac{1}{ 4}(p^2_y-p^2_z)
-iJ_3p_z+f_1,$$ 
$${\cal S}_2=J_2p_z-J_3p_y+i(J_3p_x-J_1p_z)-\frac{i}{ 2}p_yp_z+f_2,\quad
{\cal S}_3=(p_x+ip_y)^2+f_4,$$ 
$${\cal S}_4=J_3p_z+iJ_1p_y+iJ_2p_x+2J_1p_x+\frac{i}{ 4}p^2_z+f_3,\quad  
{\cal S}_5=p_z(p_x+ip_y)+f_5.$$
\end{theorem}

In \cite{KKM20052} we proved what was far from obvious, the fact that {\it no other } nondegenerate superintegrable system
separates for {\it any} special case of ellipsoidal coordinates, i.e., fixed parameter.

\begin{theorem}\label{eucgen} A 3D Euclidean nondegenerate superintegrable system admits
  separation in a special case of the generic coordinates [2111],
  [221], [23], [311], [32], [41] or [5], respectively, if and only if
  it is equivalent via a Euclidean transformation to system [I], [II], [III], [IV], [V], [VI] or [VII],
  respectively.
\end{theorem}

This does not settle the problem of classifying all 3D nondegenerate
superintegrable systems in complex Euclidean space, for we have not
excluded the possibility of such systems that separate only in
degenerate separable coordinates. In fact we have already studied two
such systems in \cite{KKM20051}:
$$[O]\quad V(x,y,z)=\alpha x+\beta y+\gamma z+\delta(x^2+y^2+z^2).
$$
\be [OO]\quad V(x,y,z)=\frac{\alpha}{2}(x^2+y^2+\frac14 z^2)+\beta x+\gamma y +\frac{\delta}{z^2}.
\ee

\section{Polynomial ideals}
In this section we introduce a very different way of studying and
classifying  superintegrable systems, through polynomial ideals. Here we
confine our analysis to 3D Euclidean superintegrable systems with
nondegenerate potentials. Thus we can set  $G\equiv 0$ in the 18
fundamental equations for the derivatives $\partial_ia^{jk}$. Due to the
linear conditions (\ref{int11}) all of the functions
$A^{ij},B^{ij},C^{ij}$ can be expressed in terms of the 10 basic
terms (\ref{10terms}).
Since the fundamental equations admit 6 linearly independent
solutions $a^{hk}$ the integrability conditions
$\partial_ia^{hk}_\ell=\partial_\ell a^{hk}_i$ for these equations must
be satisfied identically. As follows from \cite{KKM20051}, these conditions
plus the integrability conditions (\ref{int31}) for the potential
allow us to compute the 30 derivatives $\partial_\ell D^{ij}$ of the
10 basic terms (equations (\ref{diffconds}) in what follows). Each is a quadratic polynomial in the 10 terms.  In
addition there are 5 quadratic conditions remaining, equation (31) in \cite{KKM20051} with  $G\equiv 0$. 

These 5 polynomials determine an ideal $\Sigma'$. Already we see that
  the values of the 10 terms at a fixed regular point must uniquely
  determine a superintegrable system. However, choosing those values
  such that the 5 conditions $I^{(a)}$-$I^{(e)}$, listed below,  are satisfied will
  not guarantee the existence of a solution, because the conditions
  may be violated for values of $(x,y,z)$ away from the chosen regular
  point. To test this we compute the derivatives $\partial_i\Sigma'$
  and obtain a single new condition, the square of the quadratic expression $I^{(f)}$, listed below.
The polynomial $I^{(f)}$  extends the ideal.  Let $\Sigma\supset \Sigma'$ be the ideal generated by the 6
  quadratic polynomials, $I^{(a)},\cdots, I^{(f)}$:

\begin{eqnarray}\label{ideal}
I^{(a)} &=& -A^{22}B^{23} + B^{23}A^{33} + B^{12}A^{13}
     + A^{23}B^{22} - A^{12}A^{23} - A^{23}B^{33} \\
I^{(b)} &=& (A^{33})^2 + B^{12}A^{33} - A^{33}A^{22}
     - A^{12}B^{33} - A^{13}C^{33} + A^{12}B^{22}\nonumber \\
& & {}\qquad
     - B^{12}A^{22} + A^{13}B^{23} - (A^{12})^2 \nonumber\\
I^{(c)} &=& B^{23}C^{33} + B^{12}A^{33} + (B^{12})^2
     + B^{22}B^{33} - (B^{33})^2 - A^{12}B^{33} - (B^{23})^2\nonumber \\
I^{(d)} &=& -B^{12}A^{23} - A^{33}A^{23} + A^{13}B^{33}
            + A^{12}B^{23}\nonumber \\
I^{(e)} &=& -B^{23}A^{23} + C^{33}A^{23} + A^{22}B^{33}
            - A^{33}B^{33} + B^{12}A^{12} \nonumber\\
I^{(f)} &=& A^{13}C^{33} + 2A^{13}B^{23} + B^{22}B^{33}
     - (B^{33})^2 + A^{33}A^{22} - (A^{33})^2 \nonumber\\
& & {}\quad + 2A^{12}B^{22} 
     + (A^{12})^2 - 2B^{12}A^{22} + (B^{12})^2
     + B^{23}C^{33} - (B^{23})^2 - 3(A^{23})^2.\nonumber
\end{eqnarray}
It can be verified with the Gr\"obner basis package of Maple that $\partial_i\Sigma
  \subseteq \Sigma$, so that the system is closed under
  differentiation!
This leads us to a fundamental result.
\begin{theorem} Choose the 10-tuple (\ref{10terms}) at a regular
  point, such that the 6 polynomial identities (\ref{ideal}) are satisfied. Then there exists one and only one Euclidean
  superintegrable system with nondegenerate potential that takes on these values at a point.
\end{theorem}
We see that all possible nondegenerate  3D Euclidean superintegrable
systems are encoded into the 6 quadratic polynomial identities. These
identities define an algebraic  variety that generically has dimension 6, though
there are singular points, such as the origin $(0,\cdots,0)$, where the
dimension of the tangent space is greater. This result gives us the
means to classify all superintegrable systems. 

An issue is that many
different 10-tuples correspond to the same superintegrable system. How
do we sort this out? The key is that the Euclidean group E(3,{\bb C}) acts as a
transformation group on the variety and gives rise to a foliation. The
action of the translation subgroup is determined by the derivatives
$\partial_kD^{ij}$ that we have already determined \ (and will list below). The action of the
rotation subgroup on the $D^{ij}$ can be determined from the behavior
of the canonical equations (\ref{veqn1a}) under rotations. The local
action on a 10-tuple is then given by 6 Lie derivatives that are a
basis for the Euclidean Lie algebra $e(3,{\bb C})$. For ``most'' 10-tuples
${\bf D}_0$ on the 6 dimensional variety the action of the Euclidean
group is locally transitive with isotropy subgroup only the identity element. Thus the group action on such points
sweeps out a solution surface homeomorphic to the 6 parameter $E(3,{\bb C})$ itself. This
occurs for the generic Jacobi elliptic system with potential 
$$
V=\alpha(x^2+y^2+z^2)+\frac{\beta}{x^2}+\frac{\gamma}{y^2}+\frac{\delta}{z^2}.
$$
At the other extreme the isotropy subgroup of the origin
$(0,\cdots,0)$ is $E(3,C)$ itself, i.e., the point is fixed under the group
action. This corresponds to the isotropic oscillator with potential
$$ V=\alpha(x^2+y^2+z^2)+\beta x+\gamma y+\delta z.
$$
More generally, the isotropy subgroup at ${\bf D}_0$ will be $H$ and
the Euclidean  group action will
sweep out a solution surface homeomorphic to the homogeneous space
$E(3,C)/H$ and define a unique superintegrable system. For example, the isotropy subalgebra formed by the
translation and rotation generators $\{P_1,P_2,P_3,J_1+iJ_2\}$
determines a new superintegrable system $[A]$ with potential
$$ V=\alpha \left((x-iy)^3+6(x^2+y^2+z^2)\right) + \beta\left( (x-iy)^2+2(x+iy)\right)
    + \gamma (x-iy) + \delta z.
$$
Indeed, each  class of St\"ackel equivalent Euclidean
superintegrable systems is associated with a unique isotropy
subalgebra of $e(3,{\bb C})$, although not all subalgebras occur. (Indeed,
there is no isotropy subalgebra conjugate to $\{P_1,P_2,P_3\}$.)  One way to
find all superintegrable systems would be  to determine a list of all
subalgebras of $e(3,{\bb C})$, defined up to conjugacy, and then for each
subalgebra to determine if it occurs as an isotropy
subalgebra. Then we would have to resolve the degeneracy problem in which
more than one superintegrable system may correspond to a single
isotropy subalgebra.

To begin our analysis of the ideal $\Sigma$ we first determine how the rotation subalgebra $so(3,{\bb C})$ acts on the 10 variables (\ref{10terms}) and their derivatives and decompose the representation spaces into $so(3,{\bb C})$ - irreducible pieces. 
The $A^{ij}$, $B^{ij}$ and $C^{ij}$ are 10 variables that,
under the action of rotations, split into two irreducible blocks
of dimension 3 and 7.

\begin{eqnarray}
X_{+1} &=& A^{33} + 3B^{12} - 2A^{22} + i(3A^{12} + B^{33} + B^{22}) \\
X_0 &=& -\sqrt2(C^{33}+2A^{13}+B^{23}) \\
X_{-1} &=& - A^{33} - 3B^{12} + 2A^{22} + i(3A^{12} + B^{33} + B^{22}) 
\end{eqnarray}

\begin{eqnarray}
Y_{+3} &=& A^{22} + 2B^{12} + i(B^{22} - 2A^{12}) \\
Y_{+2} &=& \sqrt6(A^{13}-B^{23}+2iA^{23}) \\
Y_{+1} &=& \frac{\sqrt3}{\sqrt5} \Bigl(3A^{22} - 2B^{12} - 4A^{33}    
              + i(B^{22} - 2A^{12} - 4 B^{33})\Bigr) \\
Y_0    &=& \frac2{\sqrt5}\Bigl(2C^{33}-A^{13}-3B^{23}\Bigr) \\
Y_{-1} &=& \frac{\sqrt3}{\sqrt5} \Bigl(2B^{12} + 4A^{33} - 3A^{22}
               + i(B^{22} - 2A^{12} -4B^{33})\Bigr) \\
Y_{-2} &=& \sqrt6(A^{13}-B^{23}-2iA^{23}) \\
Y_{-3} &=& - A^{22} - 2B^{12} + i(B^{22}-2A^{12})
\end{eqnarray}

Quadratics in the variables can also be decomposed into irreducible
blocks.  There are 2 one-dimensional representations, 3 of dimension 5,
1 of dimension 7, 2 of dimension 9 and 1 of dimension 13.
\begin{eqnarray}
Z^{(1a)}_0 &=& X_0^2-2X_{-1}X_{+1} \\
Z^{(1b)}_0 &=& Y_0^2-2Y_{-1}Y_{+1}+2Y_{-2}Y_{+2}-2Y_{-3}Y_{+3} \\[5mm]
Z^{(5a)}_{\pm2} &=& X_{\pm1}^2 \\
Z^{(5a)}_{\pm1} &=& \sqrt2 X_0 X_{\pm1} \\
Z^{(5a)}_{0}  &=& \frac{\sqrt2}{\sqrt3}(X_0^2+X_{-1}X_{+1}) \\[5mm]
%Z^{(5a)}_{-1} &=& \sqrt2 X_0 X_{-1} \\
%Z^{(5a)}_{-2} &=& X_{-1}^2 \\[5mm]
%
%
Z^{(5b)}_{\pm2} &=& Y_{\pm1}^2 - \frac{\sqrt{10}}{\sqrt3} Y_0Y_{\pm2}
                     + \frac{\sqrt5}{\sqrt3} Y_{\mp1}Y_{\pm3} \\
Z^{(5b)}_{\pm1} &=& \frac{1}{\sqrt3} Y_0Y_{\pm1}
                     - \frac{\sqrt5}{\sqrt2} Y_{\mp1}Y_{\pm2}
                     + \frac{5}{\sqrt6} Y_{\mp2}Y_{\pm3} \\
Z^{(5b)}_{0}  &=& \frac{\sqrt2}{\sqrt3} Y_0^2
                     - \frac{\sqrt3}{\sqrt2} Y_{-1}Y_{+1}
                     + \frac{5}{\sqrt6} Y_{-3}Y_{+3} \\[5mm]
%Z^{(5b)}_{-1} &=&  \frac{1}{\sqrt3} Y_0Y_{-1}
%                     - \frac{\sqrt5}{\sqrt2} Y_{-2}Y_{+1}
%                     + \frac{5}{\sqrt6} Y_{-3}Y_{+2} \\
%Z^{(5b)}_{-2} &=& Y_{-1}^2 - \frac{\sqrt{10}}{\sqrt3} Y_0Y_{-2}
%                     + \frac{\sqrt5}{\sqrt3} Y_{-3}Y_{+1} \\[5mm]
%
%
Z^{(5c)}_{\pm2} &=& X_{\mp1}Y_{\pm3}
                     + \frac{1}{\sqrt{15}} X_{\pm1}Y_{\pm1}
                     - \frac{1}{\sqrt3} X_0Y_{\pm2} \\
Z^{(5c)}_{\pm1} &=& \frac{1}{\sqrt5} X_{\pm1}Y_0
                     - \frac{2\sqrt2}{\sqrt{15}} X_0Y_{\pm1}
                     + \frac{\sqrt2}{\sqrt3} X_{\mp1}Y_{\pm2} \\
Z^{(5c)}_{0}  &=& - \frac{\sqrt3}{\sqrt5} X_0Y_0
                     + \frac{\sqrt2}{\sqrt5} X_{-1}Y_{+1}
                     + \frac{\sqrt2}{\sqrt5} X_{+1}Y_{-1}
%Z^{(5c)}_{-1} &=& \frac{1}{\sqrt5} X_{-1}Y_0
%                     - \frac{2\sqrt2}{\sqrt{15}} X_0Y_{-1}
%                     + \frac{\sqrt2}{\sqrt3} X_{+1}Y_{-2} \\
%Z^{(5c)}_{-2} &=& X_{+1}Y_{-3}
%                     + \frac{1}{\sqrt{15}} X_{-1}Y_{-1}
%                     - \frac{1}{\sqrt3} X_0Y_{-2}
\end{eqnarray}

There is one 7-dimensional representation with highest weight vector
\begin{equation}
Z^{(7)}_{+3} = X_0Y_{+3} - \frac{1}{\sqrt3} X_{+1}Y_{+2}\,,
\end{equation}
two 9-dimensional representations with highest weight vectors
\begin{eqnarray}
Z^{(9a)}_{+4} &=& Y_{+2}^2 - \frac{2\sqrt3}{\sqrt5} Y_{+1}Y_{+3}  \\
Z^{(9b)}_{+4} &=& X_{+1}Y_{+3}
\end{eqnarray}
and one 13-dimensional representation
\begin{equation}
Z^{(13)}_{+3} = Y_{+3}^2\,.
\end{equation}

A linear combination of representations of the same
dimension is another representation and if we define
\begin{eqnarray}
Z_{m} &=& 2Z^{(5a)}_{m} - 5Z^{(5b)}_m + 5Z^{(5c)}_m\,,\qquad
\mbox{for $m=-2,-1,0,+1,+2$.} \\
W_0 &=& 8Z^{(1a)}_0 - 5Z^{(1b)}_0\,,
\end{eqnarray}
the algebraic variety defining the nondegenerate superintegrable
systems is given by
\begin{equation}
\label{eqn:quadidents}
Z_m=W_0=0\qquad \mbox{for $m=-2,-1,0,+1,+2$.}
\end{equation}
If $J_x$, $J_y$ and $J_z$ are Lie derivatives corresponding to
rotation about the $x$, $y$ and $z$ axes, we define
\[
J_+ = iJ_x + J_y\,, \quad J_- = iJ_x - J_y
\quad\mbox{and}\quad J_3 = iJ_z\,.
\]
then
\begin{eqnarray}
J_+f_m &=& \sqrt{(l-m)(l+m+1)}f_{m+1} \\
J_-f_m &=& \sqrt{(l+m)(l-m+1)}f_{m-1}\nonumber \\
J_3f_m &=& mf_m\nonumber
\end{eqnarray}
where $f_m$ is taken as one of $X_m$, $Y_m$, $Z_m$ or $W_0$.

Derivatives of the $X_m$ and $Y_m$ are quadratics in these
variables.  The derivatives of the $X_m$ are linear combinations
of the quadratics from the representations of dimensions 1 and 5.
In particular,
\begin{equation}\label{Xde}
\partial_iX_j \in \{2Z^{(5a)}_m+5Z^{(5b)}_m:m=0,\pm1,\pm2\}
     \cup\{Z^{(1A)}_0\}\,.
\end{equation}
Hence the quadratic identities (\ref{eqn:quadidents}) can be
used to write these derivatives as a sum of terms each of degree
at least 1 in the $X_m$.  This means that whenever all of the
$X_m$ vanish at a point, their derivatives also vanish and hence
the set $\{X_{-1},X_0,X_{+1}\}$ is a relative invariant.

The derivatives of the $Y_m$ are linear combinations
of the quadratics from the representations of dimensions 5
and 9.  
\begin{equation} 
\partial_iY_j \in \{2Z^{(5a)}_m+5Z^{(5b)}_m: -2\le m\le +2\}
     \cup \{5Z^{(9a)}_m-24Z^{(9b)}_m: -4\le m\le +4\}\,.
\end{equation}
Hence they can be written as a sum of terms each of degree
at least 1 in the $Y_m$, so\
$$\{Y_{-3},Y_{-2},Y_{-1},Y_{0},Y_{+1},Y_{+2},Y_{+3}\}$$ is a relative
invariant set.
Note that from the dimension of the spaces containing the derivatives
of the $X_m$ and $Y_m$, there must be at least 3 linear relations
among the derivatives of the $X_m$ and 7 among the derivatives
of the $Y_m$.

In a similar way we can find we can find relative invariant sets
of quadratics carrying a representation of the Lie
algebra $so(3,{\bb C})$. 
For example, the following are relative invariant sets. 
\begin{eqnarray} \label{relinv}
R_1 &=& \{X_{-1},X_0,X_{+1}\}, \\
R_2 &=& \{Y_{-3},Y_{-2},Y_{-1},Y_{0},Y_{+1},Y_{+2},Y_{+3}\},\nonumber  \\
R_3 &=& \{4Z^{(5a)}_m-15Z^{(5b)}_m:m=0,\pm1,\pm2\}\cup\{Z^{(1A)}_0\},\nonumber \\
R_4 &=& \{3Z^{(5a)}_m-5Z^{(5b)}_m:m=0,\pm1,\pm2\}\cup\{Z^{(1A)}_0\},\nonumber \\
R_5 &=& \{8Z^{(5a)}_m-5Z^{(5b)}_m:m=0,\pm1,\pm2\},\nonumber \\
R_6 &=& R_5 \cup \{5Z^{(9a)}_m+6Z^{(9b)}_m:m=0,\pm1,\pm2,\pm3,\pm4\}.\nonumber
\end{eqnarray}

Recall that the known superintegrable nondegenerate potentials are
\begin{eqnarray}\label{ndpotentials}
V_{I} &=& \alpha(x^2+y^2+z^2) + \frac\beta{x^2}
   + \frac\gamma{y^2} + \frac\delta{z^2}, \\
V_{II} &=& \alpha(x^2+y^2+z^2) + \frac{\beta(x-iy)}{(x+iy)^3}
   + \frac{\gamma}{(x+iy)^2} + \frac\delta{z^2},\nonumber \\
V_{III} &=& \alpha(x^2+y^2+z^2) + \beta{(x+iy)^2}
   + \frac{\gamma z}{(x+iy)^3}
   + \frac{\delta(x^2+y^2-3z^2)}{(x+iy)^4},\nonumber \\
V_{IV} &=& \alpha(4x^2+y^2+z^2) + \beta x
   + \frac\gamma{y^2} + \frac{\delta}{z^2},\nonumber \\
V_{V} &=& \alpha(4z^2+x^2+y^2) + \beta z
   + \frac\gamma{(x+iy)^2} + \frac{\delta(x-iy)}{(x+iy)^3},\nonumber \\
V_{VI} &=& \alpha(4x^2+4y^2+z^2-2(x-iy)^3)
   + \beta(2x+2iy-3(x-iy)^2) + \gamma(x-iy) + \frac\delta{z^2},\nonumber \\
V_{VII} &=& \alpha(x+iy) + \beta(3(x+iy)^2+z)
   + \gamma(16(x+iy)^3+x-iy+12z(x+iy))\nonumber \\
 & & {}\qquad   + \delta(5(x+iy)^4+x^2+y^2+z^2+6(x+iy)^2z),\nonumber \\
V_{O} &=& \alpha(x^2+y^2+z^2)+\beta x + \gamma y + \delta z,\nonumber \\
V_{OO} &=& \alpha(4x^2+4y^2+z^2) + \beta x + \gamma y
   + \frac\delta{z^2},\nonumber \\
V_{A} &=& \alpha((x-iy)^3+6(x^2+y^2+z^2))
   + \beta((x-iy)^2 + 2x+2iy) + \gamma(x-iy) + \delta z.\nonumber
\end{eqnarray}
The correspondence between relative invariant sets and
 potentials is in the accompanying table.

\medskip

\begin{tabular}{c|c|c|c|c|c|c|}
$V$   &$R_1$&$R_2$&$R_3$&$R_4$&$R_5$&$R_6$ \\ 
\hline
$I$   &     &     &     &     &     & \\
$II$  &     &     &     &     &     & \\
$III$ &     &     & $0$ &     &     & \\
$IV$  &     &     &     &     &     & \\
$V$   &     &     &     & $0$ &     & \\
$VI$  &     &     &     &     & $0$ & \\
$VII$ & $0$ &     & $0$ & $0$ & $0$ & \\
$O$   & $0$ & $0$ & $0$ & $0$ & $0$ & $0$ \\
$OO$  &     &     &     &     & $0$ & $0$ \\
$A$   & $0$ &     & $0$ & $0$ & $0$ & $0$
\end{tabular}

\medskip

The action of the Euclidean translation generators on the 10 basis
monomials can also be written in terms of the irreducible
representations of $so(3,{\bb C})$. (Indeed these equations are much
simpler than when written directly in terms of the
$A^{ij},B^{ij},C^{ij}$.) Using the notation 
\begin{equation}
%\partial_\pm = \partial_x\pm i\partial_y
\partial_\pm = i\partial_y\pm\partial_x
\end{equation}
\begin{equation}\label{Zde}
Z^{(5X)}_m = 5Z^{(5b)}_m + 2Z^{(5a)}_m\,, \qquad
Z^{(9Y)}_m = 24Z^{(9b)}_m - 5Z^{(9a)}\,.
\end{equation} we obtain the fundamental differential relations:
\begin{eqnarray}\label{diffconds}
\partial_-X_{+1} &=& \frac{1}{30\sqrt6}Z^{(5X)}_0 - \frac19Z^{(1A)}_0,\quad
\partial_+X_{+1} = \frac1{30}Z^{(5X)}_{+2}, \\
\partial_zX_{+1} &=& -\frac1{60}Z^{(5X)}_{+1},\quad
\partial_-X_{0}  = \frac1{30\sqrt2}Z^{(5X)}_{-1}, \nonumber\\
\partial_+X_{0}  &=& \frac1{30\sqrt2}Z^{(5X)}_{+1},\quad
\partial_zX_{0}  = -\frac1{30\sqrt3}Z^{(5X)}_0 - \frac1{9\sqrt2}Z^{(1A)}_0, \nonumber\\
\partial_-X_{-1} &=& \frac1{30}Z^{(5X)}_{-2},\quad
\partial_+X_{-1} = \frac1{30\sqrt6}Z^{(5X)}_0 -\frac19Z^{(1A)}_0,\nonumber \\
\partial_zX_{-1} &=& -\frac1{60}Z^{(5X)}_{-1}, \nonumber\\[5mm]
\partial_-Y_{+3} &=& \frac1{180\sqrt7}Z^{(9Y)}_{+2} + \frac1{35}Z^{(5X)}_{+2},\quad
\partial_+Y_{+3} = \frac1{90}Z^{(9Y)}_{+4},\label{Yde} \\
\partial_zY_{+3} &=& -\frac1{180\sqrt2}Z^{(9Y)}_{+3},\quad 
\partial_-Y_{+2} = \frac1{60\sqrt{21}}Z^{(9Y)}_{+1} 
                      + \frac{\sqrt2}{35\sqrt3}Z^{(5X)}_{+1},\nonumber \\
\partial_+Y_{+2} &=& \frac1{60\sqrt3}Z^{(9Y)}_{+3},\quad
\partial_zY_{+2} = -\frac1{30\sqrt{42}}Z^{(9Y)}_{+2} + \frac1{35\sqrt6}Z^{(5X)}_{+2}, \nonumber \\
\partial_-Y_{+1} &=& \frac1{30\sqrt{42}}Z^{(9Y)}_0 + \frac{\sqrt2}{35\sqrt5}Z^{(5X)}_0 ,\quad
\partial_+Y_{+1} = \frac1{12\sqrt{105}}Z^{(9Y)}_{+2} + \frac1{35\sqrt{15}}Z^{(5X)}_{+2},\nonumber \\
\partial_zY_{+1} &=& -\frac1{12\sqrt{210}}Z^{(9Y)}_{+1} + \frac2{35\sqrt{15}}Z^{(5X)}_{+1},\quad
\partial_-Y_{0}  = \frac1{18\sqrt{70}}Z^{(9Y)}_{-1} + \frac1{35\sqrt5}X^{(5X)}_{-1},\nonumber \\
\partial_+Y_{0}  &=&  \frac1{18\sqrt{70}}Z^{(9Y)}_{+1} + \frac1{35\sqrt5}X^{(5X)}_{+1},\quad
\partial_zY_{0}  = -\frac1{45\sqrt{14}}Z^{(9Y)}_0 + \frac{\sqrt3}{35\sqrt{10}}X^{(5X)}_0,\nonumber \\
\partial_-Y_{-1} &=& \frac1{12\sqrt{105}}Z^{(9Y)}_{-2} + \frac1{35\sqrt{15}}Z^{(5X)}_{-2},\quad
\partial_+Y_{-1} = \frac1{30\sqrt{42}}Z^{(9Y)}_0 + \frac{\sqrt2}{35\sqrt{5}}Z^{(5X)}_0 ,\nonumber\\
\partial_zY_{-1} &=& -\frac1{12\sqrt{210}}Z^{(9Y)}_{-1} +  \frac2{35\sqrt{15}}Z^{(5X)}_{-1}\quad
\partial_-Y_{-2} = \frac1{60\sqrt3}Z^{(9Y)}_{-3},\nonumber \\
\partial_+Y_{-2} &=& \frac1{60\sqrt{21}}Z^{(9Y)}_{-1} + \frac{\sqrt2}{35\sqrt3}Z^{(5X)}_{-1},\quad
\partial_zY_{-2} = -\frac1{30\sqrt{42}}Z^{(9Y)}_{-2} + \frac1{35\sqrt6}Z^{(5X)}_{-2},\nonumber \\
\partial_-Y_{-3} &=& \frac1{90}Z^{(9Y)}_{-4},\quad
\partial_+Y_{-3} = \frac1{180\sqrt7}Z^{(9Y)}_{-2} +\frac1{35}Z^{(5X)}_{-2},\nonumber \\
\partial_zY_{-3} &=& -\frac1{180\sqrt2}Z^{(9Y)}_{-3}.\nonumber \\
\end{eqnarray}

In the following table we describe each of the known superintegrable
systems in terms of variables adapted to the rotation group action.
for this it is convenient to choose the 10 constrained variables in
the form $X_i,\ i=1\ldots3$ and $Y_j,\ j=1\ldots7$ with $d_X$ and
$d_Y$, respectively,  as the number of 
independent variables
on which these variables  depend.
These are defined by
\begin{eqnarray}
X_1 &=& 2A^{13} + B^{23} + C^{33}=-\frac{X_0}{\sqrt{2}},\ 
X_2 = 2A^{22} - A^{33} - 3B^{12}=\frac{X_{-1}-X_{+1}}{2},\nonumber \\
X_3 &=& 3A^{12} + B^{33} + B^{22}=\frac{X_{-1}+X_{+1}}{2},\  Y_1 = \frac12(Y_{+3}-Y_{-3}),\nonumber  %\label{Xvariables}
\end{eqnarray}

\begin{eqnarray}\label{Yvariables}
Y_2 &=& \frac1{2i}(Y_{+3}+Y_{-3}), \  
Y_3 = \frac1{2i\sqrt6}(Y_{+2}-Y_{-2}), \
Y_4 = \frac1{2\sqrt6}(Y_{+2}+Y_{-2}),\nonumber \\
Y_5 &=& \frac{\sqrt5}{2\sqrt3}(Y_{+1}-Y_{-1}), \
Y_6 = \frac{\sqrt5}{2i\sqrt3}(Y_{+1}+Y_{-1}),\
Y_7 = \frac{\sqrt5}2Y_0.
\end{eqnarray}

\medskip

\begin{tabular}{|c|c|c|c|c|}
 & $\sum_{j=1}^3X_j^2$
 & $[X_1,X_2,X_3]$
 & $d_X$
 & $[Y_1,Y_2,Y_3,Y_4,Y_5,Y_6,Y_7]$
 \\
&&& $d_Y$&\\
\hline &&&& \\
$V_I$
 & $\frac9{x^2}+\frac9{y^2}+\frac9{z^2}$
 & $\left[-\frac3x,-\frac3y,\frac3z\right]$
 & $3$
 & $\left[\frac3x,-\frac3y,0,0,-\frac3x,-\frac3y,-\frac6z\right]$
 \\

 & 
 & 
 & $3$
 & 
  \\[5mm]
$V_{II}$
 & $\frac9{z^2}$
 & $\left[-\frac6{x+iy},-\frac{6i}{x+iy},\frac3z\right]$
 & $2$
 & $\left[-\frac{6(x-iy)}{(x+iy)^2},-\frac{6i(x-iy)}{(x+iy)^2},
        0,0,-\frac6{x+iy},-\frac{6i}{x+iy},-\frac6z \right]$
  \\

 &
 & 
 & $3$
 & 
  \\[5mm]
$V_{III}$
 & $0$
 & $\left[-\frac9{x+iy},-\frac{9i}{x+iy},0\right]$
 & $1$
 & $\left[-\frac{6(x^2+y^2-2z^2)}{(x+iy)^3},-\frac{6i(x^2+y^2-2z^2)}{(x+iy)^3},
     \frac{6iz}{(x+iy)^2},\right.$
  \\
 &
 &
 &$3$
 & $\left.\frac{6z}{(x+iy)^2},
          \frac{6}{x+iy},\frac{6i}{x+iy},0 \right]$
  \\[5mm]

$V_{IV}$
 & $\frac9{y^2}+\frac9{z^2}$
 & $\left[0,-\frac3y,\frac3z\right]$
 & $2$
 & $\left[0,-\frac3y,0,0,0,-\frac3y,-\frac6z\right]$
  \\

 & 
 & 
 & $2$
 & 
  \\[5mm]
$V_{V}$
 & $0$
 & $\left[-\frac6{x+iy},-\frac{6i}{x+iy},0\right]$
 & $1$
 & $\left[-\frac{6(x-iy)}{(x+iy)^2},-\frac{6i(x-iy)}{(x+iy)^2},0,0
          -\frac6{x+iy},-\frac{6i}{x+iy},0\right]$
 \\

 & 
 & 
 & $2$
 & 
  \\[5mm]
$V_{VI}$
 & $\frac9{z^2}$
 & $\left[0,0,\frac3z\right]$
 & $1$
 & $\left[6,-6i,0,0,0,0,-\frac6z\right]$
 \\

 & 
 & 
 & $1$
 & 
 \\[5mm]
$V_{VII}$
 & $0$
 & $[0,0,0]$
 & $0$
 & $[-48(x+iy),-48i(x+iy),12i,12,0,0,0]$
 \\

 & 
 & 
 & $1$
 & 
 \\[5mm]
$V_{O}$
 & $0$
 & $[0,0,0]$
 & $0$
 & $[0,0,0,0,0,0,0]$
  \\

 & 
 & 
 & $0$
 & 
  \\[5mm]
$V_{OO}$
 & $\frac9{z^2}$
 & $\left[0,0,\frac3z\right]$
 & $1$
 & $\left[0,0,0,0,0,0,-\frac6z\right]$
  \\

 & 
 & 
 & $1$
 & 
 \\[5mm]
$V_{A}$
 & $0$
 & $[0,0,0]$
 & $0$
 & $[-2,2i,0,0,0,0,0]$
\\

 & 
 & 
 & $0$
 & 

\end{tabular}

\bigskip

In principle one could classify all  possibilities by referring to
distinct cases exhibited in the accompanying  table. Here,
however,  we  use the preceding algebraic and differential  conditions, together with the  coordinates  in which the corresponding nondegenerate system could separate, to demonstrate that our 10 known superintegrable systems are the only ones possible.

\section{Completion of the proof}
We know that in addition to the generic superintegrable systems, the only possible superintegrable systems are those that are multiseparable in nongeneric coordinates. Our strategy is 
to consider each  nongeneric separable system in a given standard form and use the 
integrability conditions associated with the corresponding separable 
potential.  If a superintegrable system permits separation in these coordinates, then by a suitable  Euclidean transformation, we can assume the system permits separation in this standard form. This information  is then  used together with the six algebraic conditions 
$I^{(a)},\cdots, I^{(f)}$, (\ref{ideal}), to deduce all the information available from algebraic 
conditions. At that point the differential equations
(\ref{diffconds}) for the $D^{ij}$ can be solved  in a
 straight forward manner to obtain the final possible superintegrable systems. In some cases the algebraic conditions alone suffice and the differential equations are unnecessary. We proceed on a case by 
case basis.

\subsection{Cylindrical systems}
For cylindrical-type  systems the  potential splits off the $z$ 
variable , i.e.,  the potential  satisfies  $V_{13}=0,V_{23}=0$ in equations (\ref{veqn1a}).  This  implies that 
$A^{13}=B^{13}=C^{13}=0$ and $A^{23}=B^{23}=C^{23}=0$. From the equations for 
$X_i, (i=1,2,3)$ and $Y_j, (j=1,\cdots,7)$ we can  deduce that $Y_7=-2X_3$. In addition it is also easy to
conclude  that
$Y_3=Y_4=0$ and $X_1=Y_5,X_2=Y_6$. 

If we add the requirement of Cartesian 
coordinate separation then $A^{12}=B^{12}=C^{12}=0$. If $X_3=0$ we
obtain potential $V_0$. If $X_3\ne 0$ then $X_3=3/z$ If $X_1=X_2=0$
then we have potential $V_{00}$. If one of $X_1,X_2$ is not zero  this leads directly to 
potential $V_I$.

{}For separation in cylindrical coordinates  
$x=r\cos\theta ,\ 
y=r\sin\theta ,\ 
z$,
 the following conditions must apply: 
$$V_{xz}=0,\  
V_{yz}=0,$$
$$(x^2-y^2)V_{xy}+xy(V_{yy}-V_{xx})+3xV_y-3yV_x=0.$$
The last condition is equivalent to  
$\partial _\theta (r\partial _r(r^2V))=0$
where $r^2=x^2+y^2$. Solving  the algebraic conditions that result, we 
determine that 
$$X_1=Y_5=-G(1+\frac{y^2}{ x^2})-\frac{3}{ x},\
X_2=Y_6=G(\frac{x}{ y}+\frac{y}{ x})-\frac{3}{ y},$$ 
$$Y_1=G(-3+\frac{y^2}{x^2})+\frac{3}{ x},\  
Y_2=G({x\over y}-3{y\over x})-{3\over y},\ 
Y_3=Y_4=0,$$
where $G$ is an unknown function. In addition we deduce that
$Y_7=-2X_3$. It is 
then easy  to show from the differential equations  that 
$X_3=\frac{3}{ z}$ or $ 0$ and that $G=0$. We  conclude that separation of this 
type occurs in cases $V_I$ and $V_{IV}$.

For  parabolic cylinder coordinates  
$x=\frac{1}{ 2}(\xi ^2-\eta ^2),\ 
y=\xi \eta ,\ 
z$, 
the conditions on the potential have the form 
$$V_{xz}=0 ,\ V_{yz}=0\ 
2xV_{xy}+y(V_{yy}-V_{xx})+3V_y=0.$$
This implies that  
$$X_1=-2F,\ 
X_2=2\frac{x}{ y}F-\frac{3}{ y},\
X_3=-C,$$ 
$$Y_1=-2F,\
Y_2=2\frac{x}{ y}F-\frac{3}{ y},\ 
Y_3=Y_4=0,
$$ 
$$Y_5=-2F,\ 
Y_6=2\frac{x}{ y}F-\frac{3}{ y} ,\ 
Y_7=2C.$$
The remaining differential equations  require that $F=0$ and 
$C={3\over z}$. This type occurs in case $V_{IV}$.

{}For  elliptic cylinder coordinates 
$x=\cosh A \cos B, y=\sinh A \sin B, z,$  the integrability conditions for the 
potential have the form 
$$V_{zx}=0 ,\ V_{yz}=0,\ 
(x^2-y^2-1)V_{xy}+xy(V_{yy}-V_{xx})+3(xV_y-yV_x)=0.$$
This and the algebraic  conditions imply  
$$X_1=(\frac{x}{ y}+\frac{y}{ x}+\frac{1}{ xy})G-\frac{3}{ x},\
X_2=(-1-\frac{x^2}{ y^2}+\frac{1}{ y^2})G-\frac{3}{ y},\ 
X_3=-C,$$ 
$$Y_1=(3\frac{x}{ y}-\frac{y}{ x}-\frac{1}{ xy})G+\frac{3}{ x},\ 
Y_2=(-\frac{x^2}{ y^2}+3+\frac{1}{ y^2})G-\frac{3}{ y},\
Y_3=Y_4=0,$$ 
$$Y_5=(\frac{x}{ y}+\frac{y}{ x}+\frac{1}{ xy})G-\frac{3}{ x},\
Y_6=(-1-\frac{x^2}{ y^2}+\frac{1}{ y^2})G-\frac{3}{ y},\ 
Y_7=2C.$$
The remaining differential equations require  $G=0$, and $C=-\frac{3}{ z}$ or $0$
corresponding to systems  $V_I$ and $V_{IV}$.

In semihyperbolic coordinates $x+iy=4i(u+v),x-iy=2i(u-v)^2$ the extra 
integrability condition is 
$$(1+ix+y)(V_{xx}-V_{yy})+2(-2i-x+iy)V_{xy}+3iV_x-3V_y=0.$$
The algebraic conditions  yield the requirements  
$$X_1=Y_5=G,\  
X_2=-G,\ 
X_3=-C,\ Y_3=Y_4=0,\ $$
$$Y_1=\frac{3}{ 2}i+\frac{i}{ 2}(x-iy)G,\
Y_2=-\frac{3}{ 2}+\frac{1}{ 2}(-x+iy)G,\
Y_6=iG,\ 
Y_7=2C.$$
This leads to potentials $V_A$ and $V_{VI}$.

{}For hyperbolic coordinates $x+iy=rs,x-iy= 
(r^2+s^2)/ rs, z$,  the integrability condition is 
$$(1+ixy)(V_{yy}-V_{xx})+i(x^2-y^2-2)V_{xy}+3i(xV_y-yV_x)=0.$$
The algebraic conditions  imply $Y_7=2X_3=2C$ and 
$$X_1=Y_5=(xy-iy^2-2i)G-\frac{6}{ x+iy},\ 
X_2=Y_6=-(x^2-ixy-2)G-\frac{6i}{ x+iy},\ 
$$
$$Y_1= \frac{3yx^2-2ix-y^3-2y}{ x+iy}G - \frac{6(x-iy)}{ (x+iy)^2},\ 
Y_2=- \frac{x^3-3xy^2-2x+2iy}{ x+iy}G - i\frac{6(x-iy)}{ (x+iy)^2}.$$
This yields potential $V_{II}$.

\subsection{Radial-type coordinates}
We consider  systems  that have a radial coordinate $r$ as one of the
separable coordinates. The two other coordinates are separable 
 on the complex two dimensional sphere. We  first consider 
spherical coordinates  
$x=r\sin\theta \cos\varphi$,
$y=r\sin\theta \sin\varphi$,
$z=r\cos\theta$. 
The integrability conditions on the potential have the form 
$$(x^2-y^2)V_{xy}+xzV_{yz}-yzV_{xz}+xy(V_{yy}-V_{xx})+3xV_y-3yV_x=0,$$ 
$$(x^2-z^2)V_{xz}+xz(V_{zz}-V_{xx})+xyV_{yz}-zyV_{xy}+3xV_z-3zV_x=0,$$ 
$$(y^2-z^2)V_{yz}+yz(V_{zz}-V_{yy})+xyV_{xz}-zxV_{xy}+3yV_z-3zV_y=0,$$
$$xV_{yz}-yV_{xz}=0.$$
Note that the first three conditions are not independent and only two are 
required. For any potential that separates in spherical coordinates, one 
additional condition is required. Indeed, 
 if $r,u$ and $v$ are any form of separable spherical-type coordinates then
the potential must have the functional form 
\be\label{star1}V=f(r)+g(u,v)/r^2,\ee
it being understood that $u$ and $v$ are coordinates on the complex two 
dimensional sphere and $r$ is the radius. It is then clear that 
$r^2V=r^2f(r)+g(u,v)$. As a consequence there are the conditions  
$ \partial _r\partial _\lambda (r^2V)=0$ ,
where $\lambda =u,v$. Noting that 
$$x\partial _xF+y\partial _yF+z\partial _zF=DF=r\partial _rF$$
 and that  
$$J_1F=y\partial _zF-z\partial _yF=a(u,v)\partial _uF+b(u,v)\partial _vF,$$
with similar expressions for $J_2F$ and $J_3F$,  we conclude that the conditions 
(\ref{star1}) are equivalent to any two of the three conditions 
${1\over r^2}J_iD(r^2V)=0$.
These are indeed the three conditions we have given. If we now solve all the 
algebraic conditions, we determine that 

$$X_1=Y_5=-\frac{(x^2+y^2)}{ xy} G -\frac{3}{ x},\  
X_2=\frac{(x^2+y^2)}{ y^2} G -\frac{3}{y},\
X_3=\frac{3}{ z},\ Y_7=-\frac{6}{z},
$$
$$Y_1=- \frac{3x^2-y^2}{xy} G +\frac{3}{ x},\ 
Y_2= \frac{x^2-3y^2}{ y^2} G -\frac{3}{ y},\ Y_3=Y_4=0.
$$
{}From this we see that the remaining differential equations give $G=0$
and  we obtain solution $V_I$.

We now consider  horospherical 
coordinates on a complex 2 sphere. viz. 
$$x+iy=-i\frac{r}{ v}(u^2+v^2),\ 
x-iy=i\frac{r}{v},\ 
z=-ir\frac{u}{ v}.
$$
The extra integrability condition in this case is  
$$z(V_{xx}-V_{yy})+2izV_{xy}-(x+iy)(V_{xz}+iV_{yz})=0.$$
Solving the algebraic conditions we conclude that 
$$X_1=iX_2=\frac{(x+iy)}{ z} G - \frac{6}{ x+iy},\
X_3= \frac{(x+iy)^2}{ z^2}G +\frac{3}{z},
$$
$$Y_1=iY_2=-4 \frac{z}{ (x+iy)} G -6 \frac{(x-iy)}{ (x+iy)^2},\ 
Y_3=iY_4=-2iG,
$$
$$Y_5=iY_6=-4 \frac{(x+iy)}{ z} G - \frac{6}{ (x+iy)},\ 
Y_7=-2 \frac{(x+iy)^2}{ z^2}G -\frac{6}{ z}.
$$
The derivative conditions give $G=0$, so this corresponds to solution $V_{II}$.

Conical coordinates are also radial-type: 
$$x^2=r^2\frac{(u-e_1)(v-e_1)}{ (e_1-e_2)(e_1-e_3)},\
y^2=r^2\frac{(u-e_2)(v-e_2)}{ (e_2-e_1)(e_2-e_3)},
$$
$$z^2=r^2\frac{(u-e_3)(v-e_3)}{ (e_3-e_2)(e_3-e_1)}.
$$
The  extra integrability condition is  
$$3(e_2-e_3)yzV_x+3(e_3-e_1)xzV_y+3(e_1-e_2)xyV_z+ 
xyz[(e_2-e_3)V_{xx}+(e_3-e_1)V_{yy}+(e_1-e_2)V_{zz}]$$
$$+z[(e_3-e_1)y^2+(e_2-e_3)x^2+(e_2-e_1)z^2]V_{xy} 
+y[(e_1-e_2)z^2+(e_2-e_3)x^2+(e_1-e_3)y^2]V_{xz}$$ 
$$+x[(e_1-e_2)z^2+(e_3-e_2)x^2+(e_1-e_3)y^2]V_{yz}=0.$$
The algebraic conditions yield immediately solution $V_I$ with  
$$X_1=-\frac{3}{ x},\  X_2=-\frac{3}{ y},\
X_3=\frac{3}{ z},\ Y_1=\frac{3}{ x},\  
Y_2=-\frac{3}{ y},$$
$$Y_3=Y_4=0,\
Y_5=-\frac{3}{ x},\
Y_6=-\frac{3}{ y},\
Y_6?=-\frac{6}{ x}.$$

For degenerate type elliptic polar coordinates (type 1) we can write 
$$x+iy = \frac{r}{ \cosh A\cosh B},\ 
2x =r[\frac{\cosh A}{ \cosh B} + \frac{\sinh B}{ \sinh A}],\ 
z=r\tanh A\tanh B.
$$
The extra integrability condition is 
$$3(x+iy)^2V_z-3xzV_x-3i(2x+iy)zV_y-2i(x+iy)(z^2+ixy)V_{yz}-2(y^2+z^2)(x+iy)V_{xz}
$$
$$+2iz(z^2+y^2)V_{xy}+z(x+iy)^2V_{zz}+z(z^2+y^2)V_{xx}-z(x^2+z^2+2ixy)V_{yy}=0.
$$
Solving the algebraic conditions we deduce that 
$$X_1=-\frac{2}{ x}(y-ix)G-\frac{6}{ x+iy},\ 
X_2=-i\frac{2}{ x}(y^2-x^2+z^2-ixy)(y-ix)G-\frac{6i}{ x+iy},
$$
$$X_3=-\frac{2i}{ xz}(z^2+y^2-ixy)(y-ix)^2G+\frac{3}{ z},\
Y_1=-\frac{1}{ x}(-y^3+3x^2y+2z^2y-6iz^2x)G-6\frac{(x-iy)}{(x+iy)^2},
 $$
$$Y_2=-\frac{i}{ x}(-3ixy^2+ix^3+2z^2y-6iz^2x)G-6i\frac{x-iy}{ (x+iy)^2},\
Y_3=iY_4=2z \frac{(y-ix)^2}{ x}G,$$
$$
Y_5=-\frac{3}{ x}(-3y^2+5ixy+2z^2)(y-ix)G-\frac{6}{ x+iy},
$$
$$ Y_6=-\frac{i}{ 6}(-8y^2+13ixy+3x^2+2z^2)(y-ix)G-\frac{6i}{ x+iy},\ 
Y_7=-\frac{2i}{ xz}(-2y^2+2ixy+3z^2)(y-ix)^2G-\frac{6}{ z}.
$$
The differential conditions require $G=0$, leading to a 
type $V_{II}$ potential.

For degenerate elliptic coordinates (type 2) on the complex 2 
sphere we have
$$x+iy=ruv,\  
x-iy=\frac{1}{ 4}r{(u^2+v^2)^2}{ u^3v^3},\
z=-\frac{i}{ 2}r \frac{u^2-v^2}{ uv}.
$$
The corresponding integrability condition is 
$$3(z^2+ixy-y^2)V_x+3i(z^2-x^2-ixy)V_y-3iz(y-ix)V_z$$
$$-i(-ixy^2+y^3+iz^2x+yz^2)V_{xx}+i(ix^3-x^2y+iz^2x+tz^2)V_{xx}$$ 
$$+i(-ix+y)(x^2+y^2)V_{zz}+2(x^2y+yz^2+ixy^2+ixz^2)V_{xy}$$ 
$$-2iz(x^2+y^2)V_{yz}-2z(x^2+y^2)V_{xz}=0.$$
The solutions to the algebraic conditions are 
$$X_1=-2iz(ix+2y)(y-ix)G-\frac{9}{ x+iy},\
X_2=2z(y+ix)(y-ix)G-\frac{9i}{ x+iy},\
$$
$$X_3=2(-ix+y)(x^2+y^2)G,\ Y_3=2i(yz^2+iz^2x-ixy^2-x^2y)G+ \frac{6iz}{ (x+iy)^2},
$$
$$Y_1=-iY_2=i\frac{(-3y^2-3x^2+4z^2)z(ix+y)}{ ix-y} G + 6 
\frac{(-x^2-y^2+2z^2)}{ (x+iy)^3},
$$ 
$$Y_4=(2yz^2+2iz^2x-y^3+ixy^2+x^2y-ix^3)G+\frac{6}{ (x+iy)^2},\ 
Y_5=iz(y+3ix)(y-ix)G+\frac{6}{ x+iy},$$
$$
Y_6=-z(ix+3y)(y-ix)G+\frac{6i}{ x+iy},\
Y_7=(-ix+y)(x^2+y^2)G.$$
The differential conditions hold only if $G=0$. This is system $V_{III}$. 

\subsection{Spheroidal coordinates}
 We take these as 
$$x=\sinh A\cos B\cos\varphi,\ 
y=\sinh A\cos B\sin\varphi,\
z=\cosh A\sin B.$$
The integrability conditions for the potential are  
$$-3zV_x+3xV_z+zx(V_{zz}-V_{xx})-zyV_{xy}+(1+x^2+y^2-z^2)V_{zx}=0,$$
$$-3zV_y+3yV_z+zy(V_{zz}-V_{yy})-zxV_{xy}+(1+x^2+y^2-z^2)V_{zy}=0,$$ 
$$yV_{zx}-xV_{zy}=0.$$
The solutions of the algebraic conditions are  
$$X_1=Y_5=-\frac{y}{x}(x^2+y^2)G-\frac{3}{ x},\
X_2=Y_6=(x^2+y^2)G-\frac{3}{ y},\
X_3=\frac{3}{ z},
$$
$$Y_1=-\frac{y}{ x}(-y^2+3x^2)G+\frac{3}{r x},\
Y_2=(-3y^2+x^2)G-\frac{3}{ y},\
Y_7=-\frac{6}{ z}.$$
From the differential conditions we see that  $G=0$, and  obtain potential $V_I$.

\subsection{Horospherical coordinates}
These are
$$x+iy=\sqrt{\rho \nu },\ x-iy= 4 \frac{\rho +\nu -\rho \nu \mu }{ \sqrt{\rho \nu }},\
z=2\sqrt{\rho \nu \mu }.$$
The corresponding integrability conditions for the potential are 
$$(x^2-ixy-z^2)V_{zx}+(yx-iy^2+iz^2)V_{zy}+i(x+iy)zV_{xy}+zx(V_{zz}-V_{xx})+izy(
V_{yy}-V_{zz})=0,$$ 
$$(x^2-y^2)V_{xy}+xy(V_{yy}-V_{xx})+zxV_{zy}-yzV_{zx}-3yV_x+3xV_y=0,
$$
$$z(V_{xx}-V_{yy})-2izV_{xy}+(ix+y)V_{zy}+(-x+iy)V_{zx}=0.$$
The solutions to all the algebraic conditions are 
$$X_1=-iX_2=- \frac{i(x+iy)}{ z} G - \frac{6}{x+iy},\
X_3= \frac{i(x+iy)^2}{ z^2}G+\frac{3}{ z},
$$
$$Y_1=-iY_2=- \frac{4iz}{ x+iy} G - 6 \frac{x-iy}{ (x+iy)^2},\
Y_3=iY_4=2G,
$$ 
$$Y_5=-iY_6=4\frac{i(x+iy)}{ z} G - \frac{6}{ x+iy},\ 
Y_7=-2i \frac{(x+iy)^2}{ z^2} G - \frac{6}{ z}.$$
The differential conditions require  $G=0$ and this gives potential $V_{II}$.

\subsection{Rotational parabolic coordinates}
For these coordinates
$x=\xi \eta \cos\varphi $ ,  $y=\xi \eta \sin\varphi$, $ z=\frac{1}{ 2}(\xi ^2-\eta ^2)$.
The required conditions on the potential are
$$xy(V_{yy}-V_{xx})+(x^2-y^2)V_{xy}-yzV_{yz}+xzV_{xz}-3yV_x+3xV_y=0,$$ 
$$x^2(V_{xx}-V_{zz})+y^2(V_{yy}-V_{zz})+2xyV_{xy}+2zxV_{yz}+2xzV_{zx}+3xV_x+3yV_y=0,$$
$$xV_{zy}-yV_{zx}=0.$$
These integrability conditions directly produce the solution 
$$X_1=-\frac{3}{ x} , \
X_2=-\frac{3}{y} , \
X_3=0 ,\
Y_1=\frac{3}{x} , \
Y_2=-\frac{3}{ y} ,
$$
$$Y_3=Y_4=0,\
Y_5=-\frac{3}{ x} , \
Y_6=-\frac{3}{ y} , \
Y_7=0.
$$
This is a permuted version of potential $V_{IV}$.

We have covered all possibilities for separable coordinates and found exactly which superintegrable system separates in each coordinate system It follows that our list of 10 superintegrable systems is complete. Another interesting consequence of this analysis is 
\begin{theorem} For every orthogonal separable coordinate system there is at least one nondegenerate superintegrable system  that separates in these coordinates.
\end{theorem}
On the other hand,   no nondegenerate superintegrable system permits
separation in  nonorthogonal heat-type coordinates. Potential
$V_{VII}$ is the only generic system that separates in generic
coordinates alone. 

\section{Outlook}   The basic structure and classification problems
for 2D second order superintegrable systems have been solved,\cite{KKMP,KKM20042,KMJP2,KMJP3,DASK2005}. For 3D
systems the  corresponding problems are much more complicated, but we
have now  achieved a verifiably complete
classification of the possible nondegenerate potentials in 3D
Euclidean space. There are 10 such potentials, as compared to 11 in
2D. To finish the classification of nondegenerate potentials for all 3D conformally flat spaces the main task remaining is the classification on the 3-sphere, probably not difficult. This is because all conformally flat systems can be obtained from flat space and the 3-sphere by St\"ackel transforms. The new idea used here that made the complete verifiable classification practical was the association of nondegenerate superintegrable systems with points on an algebraic variety on which the Euclidean group acts to produce foliations. In the future we hope to refine this approach to give a direct classification using only the algebraic variety and group action. Here we had also to rely on basic results from separation of variables theory  to simplify the calculations. In distinction to the 2D case, which is special, the 3D classification problem seems to have all of the ingredients that go into the corresponding nondegenerate potential classification problem in  $n$ dimensions.  The number of nondegenerate potentials grows rapidly with dimension: the number of generic potentials alone is $\sum_{j=0}^np(j)$, where $p(j)$ is the number of partitions of $j$. The algebraic variety approach should be generalizable to this case.

Nondegenerate potentials for 3D superintegrable systems are just the most symmetric. There is also ``fine structure,"  a hierarchy of various classes of degenerate potentials with fewer than 4 parameters. The structure and classification theory for these systems has just gotten underway, with initial results for 3 parameter FLI systems. \cite{KKM20071}. Sometimes a quadratic algebra structure exists and sometimes it does not. Extension of these methods to complete the fine structure analysis for 3D systems
appears relatively straightforward.  The  analysis
can be extended to  2 parameter and 1 parameter potentials
with 5 functionally linearly independent second order symmetries. Here
first order PDEs for the potential appear as well as second order,
and Killing vectors may occur. Another  class of 3D superintegrable
systems is that for which the 5 functionally independent symmetries are
functionally linearly dependent. This class is related to the Calogero
potential \cite{CAL1,WOJW,HMS} and necessarily leads to first order PDEs for the
potential, as well as second order \cite{KKM20061}. However, the
integrability 
methods discussed here should be able to handle this class with no
special difficulties. On a deeper level, we think that the algebraic
geometry approach can be extended to determine the
possible superintegrable systems in all these cases.

Finally, the algebraic geometry related results that we have described
in this paper suggest strongly that there is an underlying
geometric structure to superintegrable systems that is not apparent
from the usual presentations of these systems.

\medskip
\noindent 
{\bf Acknowledgement}:  The authors wish to thank Thomas Wolf and Greg Reid for  very helpful consultations  on Gr\"obner basis techniques and on  numerical methods in the study of algebraic varieties.


\begin{thebibliography}{99}
%-----------------------------------------------------------------------
\bibitem{WOJ}
Wojciechowski S.,
Superintegrability of the Calogero-Moser System.
{\it Phys. Lett.},  {\bf V. A 95},  279--281, (1983).
%-----------------------------------------------------------------------
\bibitem{EVA}Evans N.W.,
Superintegrability in Classical Mechanics;
{\it
Phys. Rev.}\  {\bf V. A 41},  5666--5676 (1990); Group Theory of the
Smorodinsky-Winternitz System; {\it J. Math. Phys.}\   {\bf V. 32},  3369, (1991)
%-----------------------------------------------------------------------
\bibitem{EVAN}
Evans N.W.,
Super-Integrability of the Winternitz System;
{\it Phys. Lett.}\  {\bf V.A 147},  483--486, (1990).
%-----------------------------------------------------------------------
\bibitem{FMSUW}
Fri\v s J., Mandrosov V., Smorodinsky Ya.A,  Uhl\'\i r  M. and Winternitz P.,
On Higher Symmetries in Quantum Mechanics; 
{\it Phys. Lett.}\   {\bf V.16},  354--356, (1965).
%-----------------------------------------------------------------------
\bibitem{FSUW}
Fri\v s J.,  Smorodinskii Ya.A.,  Uhl\'\i r M. and Winternitz P., 
Symmetry Groups in Classical and Quantum Mechanics; {\it
Sov. J. Nucl. Phys.}\  {\bf V.4},  444--450, (1967).
%-----------------------------------------------------------------------
\bibitem{MSVW}
Makarov A.A.,  Smorodinsky Ya.A., Valiev Kh. and
Winternitz P., 
A Systematic Search for Nonrelativistic Systems with
Dynamical Symmetries.
{\it Nuovo Cimento}, {\bf V. 52},  1061--1084, (1967).
%-----------------------------------------------------------------------
\bibitem{CALO} Calogero F.,
Solution of a Three-Body Problem in One
Dimension.
{\it J. Math. Phys.}\  {\bf V.10},  2191--2196, (1969).
%-----------------------------------------------------------------------

\bibitem{CIMC}
Cisneros A. and McIntosh H.V., 
Symmetry of the
Two-Dimensional Hydrogen Atom.
{\it J. Math. Phys.}\  {\bf V.10},  277--286, (1969).
%-----------------------------------------------------------------------

\bibitem{KKM20061}
Kalnins E.G., Kress J.M, and  Miller W.Jr.,  
Second  order superintegrable systems in conformally
flat spaces.  V: 2D and 3D quantum systems. {\it
  J. Math. Phys.},  {\bf V.47}, 093501, (2006).
%-----------------------------------------------------------------------
\bibitem{MPSTAN}
L.~G.~Mardoyan, G.~S.~Pogosyan, A.~N.~Sissakian and
V.~M.~Ter-Antonyan.
Elliptic Basis for a Circular Oscillator.
{\it Nuovo
Cimento},\ {\bf B 88},  43 (1985).
%-----------------------------------------------------------------------
Two-Dimensional Hydrogen Atom: I.\ Elliptic Bases;
{\it
Theor.~Math.~Phys.}\ {\bf 61},  1021 (1984);
%-----------------------------------------------------------------------
Hidden symmetry, Separation of Variables and Interbasis Expansions in
the Two-Dimensional Hydrogen Atom.
{\it J. Phys.},\ {\bf A 18}, 455
(1985).
%-----------------------------------------------------------------------


\bibitem{GZLU}
Ya.~A.~Granovsky, A.~S.~Zhedanov and I.~M.~Lutzenko.
Quadratic Algebra as a `Hidden' Symmetry of the Hartmann Potential;
{\it J. Phys.}\ {\bf A 24},  3887 (1991).
%-----------------------------------------------------------------------

\bibitem{BDK}
Bonatos D., Daskaloyannis C. and Kokkotas K., 
Deformed
Oscillator Algebras for Two-Dimensional Quantum Superintegrable
Systems;
{\it Phys. Rev.},  {\bf  V.A 50},  3700--3709, (1994).
%-----------------------------------------------------------------------


\bibitem{GPS}
Grosche C., Pogosyan G.S., Sissakian A.N.,
Path Integral
Discussion for Smorodinsky - Winternitz Potentials:I. Two - and Three
Dimensional Euclidean Space.
{\it Fortschritte der Physik}, {\bf  V. 43},
453--521, (1995).
%-----------------------------------------------------------------------
\bibitem{KKMP}
Kalnins E.G., Kress J.M., Miller W.Jr. and Pogosyan G.S.,
{\it Completeness of superintegrability in two-dimensional constant
curvature spaces.}
{\it J.~Phys.~A:~Math~Gen.},\  {\bf V.34}, 4705--4720, (2001).
% 4705--4720
%-----------------------------------------------------------------------
\bibitem{KKW}
Kalnins E.G., Kress J.MN., and Winternitz P.,
{\it Superintegrability in a two-dimensional space of non-constant curvature.}
{\it J.~Math.~Phys.}\  {\bf V.43}, 970--983, (2002).
% 970--983.
%-----------------------------------------------------------------------
\bibitem{KKMW}
Kalnins E.G., Kress J.M., Miller, W.Jr. and Winternitz P.,
{\it Superintegrable systems in Darboux spaces.}
{\it J.~Math.~Phys.},\  {\bf  V.44}, 5811--5848, (2003).
% 
%-----------------------------------------------------------------------
\bibitem{RAN}  
Ra\~nada M.F.,
Superintegrable $n$=2 systems, quadratic constants of motion, and
potentials of Drach. 
{\it J. Math. Phys.}, {\bf V.38}, 4165--78, (1997).
%-----------------------------------------------------------------------
\bibitem{KMWP} 
Kalnins E.G., Miller W.Jr.,  Williams G.C.  and Pogosyan G.S.,
On superintegrable symmetry-breaking potentials in
  $n$-dimensional Euclidean space.
{\it J.~Phys.~A:~Math~Gen.},\  {\bf  V.35}, 4655--4720, (2002).
% 4705--4720
%-----------------------------------------------------------------------

\bibitem{KKM20051}
Kalnins E.G., Kress J.M, and  Miller W.Jr., 
Second  order superintegrable systems in conformally
flat spaces.  III: 3D  classical structure theory. {\it
  J. Math. Phys.}, {\bf V.46}, 103507, (2005).
%-----------------------------------------------------------------------

\bibitem{KKM20071}
Kalnins E.G., Kress J.M, and  Miller W.Jr.,  Fine structure for second
order superintegrable systems   (to appear in the procedings volumes
of the 
IMA program on "Symmetries and Overdetermined Systems of Partial 
Differential Equations." ) 2007.
%-----------------------------------------------------------------------

\bibitem{KKM20052}
Kalnins E.G., Kress J.M, and  Miller W.Jr.,  
Second  order superintegrable systems in conformally
flat spaces.  IV: The classical 3D St\"ackel transform and 3D
classification theory. {\it
  J. Math. Phys.}, {\bf  V.47}, 043514, (2006).
%-----------------------------------------------------------------------
\bibitem{ERNIE}
Kalnins E.G.,
{\it Separation of Variables for Riemannian Spaces of Constant
Curvature}, Pitman, Monographs and Surveys in Pure and Applied Mathematics
V.28, 184--208,
 Longman, Essex, England,
 1986. 
%----------------------------------------------------------------------- 
\bibitem{MIL88} Miller W.Jr., 
Mechanisms for variable separation in partial
differential equations and their relationship to group theory. In
{\it Symmetries and Non-linear Phenomena} pp.~188--221, World Scientific,
1988.
%-----------------------------------------------------------------------


\bibitem{KKWMPOG}
E.G.Kalnins, J.M.Kress, W.Miller Jr. and G.S.Pogosyan.
Non degenerate superintegrable systems in $n$ dimensional complex Euclidean
spaces. (To appear in 
Physics of Atomic Nuclei)
%-----------------------------------------------------------------------

\bibitem{KKM20041}
Kalnins E.G., Kress J.M, and  Miller W.Jr., 
Second  order superintegrable systems in conformally
flat spaces.  I: 2D classical structure theory. {\it J. Math. Phys.}, {\bf
V.46}, 053509 (2005).
%-----------------------------------------------------------------------



\bibitem{Bocher}M. B\^ocher. \"Uber die Riehenentwickelungen der Potentialtheory. 
Symmetry and Separation of Variables.
{\it Teubner},\ Leipzig, 1894.
%-----------------------------------------------------------------------

\bibitem{KMR}
E.~G.~Kalnins, W.~Miller and G.~K.~Reid,
Separation of variables for Riemannian spaces of constant
curvature. I. Orthogonal separable coordinates for $S_c$ and $E_{nC}$.
{\it Proc R. Soc. Lond. A}\ {\bf 39}, 183--206 (1984).
%-----------------------------------------------------------------------


\bibitem{KKM20042}
Kalnins E.G., Kress J.M, and  Miller W.Jr., 
Second  order superintegrable systems in conformally
flat spaces.  II: The classical 2D St\"ackel transform. {\it J. Math. Phys.},
{\bf V.46}, 053510, (2005).
%-----------------------------------------------------------------------
\bibitem{KMJP2}  
E.~G.~Kalnins, W.~Miller,~Jr. and G.~S.~Pogosyan.
Completeness of multiseparable superintegrability in $E_{2,C}$.
{\it J.~Phys.~A:~Math~Gen.}\ {\bf 33}, 4105 (2000).
%-----------------------------------------------------------------------
\bibitem{KMJP3}  
E.~G.~Kalnins, W.~Miller Jr. and G.~S.~Pogosyan.
Completeness of multiseparable superintegrability on the complex 2-sphere.
{\it J.~Phys.~A:~Math~Gen.}\ {\bf 33}, 6791-6806 (2000).
%-----------------------------------------------------------------------

\bibitem{DASK2005}
C.~ Daskaloyannis and K~Ypsilantis.          Unified treatment and classification of  superintegrable
systems with integrals quadratic in momenta
on a two dimensional manifold. {\it J. Math. Phys.}\ {\bf 47}, 042904  (2006).     
%-----------------------------------------------------------------------





\bibitem{CAL1}
F.~Calogero.
Solution to the one-dimensional $N$-body problems with quadratic
and/or inversely quadratic pair potentials.
{\it J.  Math. Phys.} {\bf  12},  419-436, (1971).
%-----------------------------------------------------------------------
\bibitem{WOJW}
S.~ Rauch-Wojciechowski and C.~Waksj\"o.
What an effective criterion of separability says about the Calogero
type systems.
{\it J. Nonlinear Math. Phys.} {\bf  12}, Suppl. 1  535-547, (2005).
%-----------------------------------------------------------------------
\bibitem{HMS}
J.\ T.\ Horwood, R.\ G.\  McLenaghan and R.\  G.\ Smirnov.
Invariant classification of orthogonally separable Hamiltonian systems
in Euclidean space.
{\it Comm.  Math. Phys.} {\bf  259},  679-709, (2005);
%-----------------------------------------------------------------------

\end{thebibliography}
\end{document}